\documentclass[twocolumn, trackchanges]{aastex62}

\usepackage[normalem]{ulem}
\usepackage{color}

\newcommand\redsout{\bgroup\markoverwith{\textcolor{red}{\rule[0.5ex]{2pt}{1.0pt}}}\ULon}

\usepackage{times}
\usepackage{graphicx}
\usepackage{amssymb}
\usepackage{amsmath}
\usepackage{pifont}
\usepackage{graphics}
\usepackage{xcolor}
\usepackage{epsfig}
\usepackage{verbatim}
\usepackage{booktabs}
\usepackage{apjfonts}

\newcommand{\Mstar}{M\textsubscript{$\star$}} 
\newcommand{\Rstar}{R\textsubscript{$\star$}} 

\newcommand{\bjdtdb}{\ensuremath{\rm {BJD_{TDB}}}}
\newcommand{\feh}{\ensuremath{\left[{\rm Fe}/{\rm H}\right]}}

\newcommand{\teff}{\ensuremath{T_{\rm eff}\,}}

\newcommand{\msun}{\ensuremath{\,M_\Sun}}
\newcommand{\rsun}{\ensuremath{\,R_\Sun}}
\newcommand{\lsun}{\ensuremath{\,L_\Sun}}

\newcommand{\mj}{\ensuremath{\,M_{\rm J}}}
\newcommand{\rj}{\ensuremath{\,R_{\rm J}}}
\newcommand{\fave}{\langle F \rangle}
\newcommand{\fluxcgs}{10$^9$ erg s$^{-1}$ cm$^{-2}$}

\newcommand{\kms}{\,km\,s$^{-1}$}
\newcommand{\loggp}{\ensuremath{\log{g_{\rm P}}}}
\newcommand{\loggstar}{\ensuremath{\log{g_\star}}}
\newcommand{\teq}{\ensuremath{T_{\rm eq}}}
\newcommand{\vsini}{\ensuremath{v\sin{I_*}}}

\newcommand{\mstarval}{0.945^{+0.060}_{-0.054}}
\newcommand{\rstarval}{0.995\pm0.015}
\newcommand{\lstarval}{1.082^{+0.051}_{-0.048}}

\newcommand{\loggstarval}{4.418^{+0.026}_{-0.025}}
\newcommand{\teffstarval}{5900\pm49}
\newcommand{\fehstarval}{-0.105\pm0.077}
\newcommand{\fehstarvalshort}{-0.105}
\newcommand{\agestarval}{6.3^{+3.5}_{-3.2}}

\newcommand{\diststarval}{125.42\pm0.35}

\newcommand{\vsinistarval}{2.42\pm0.50}

\newcommand{\periodval}{2.255353^{+0.000031}_{-0.000030}}
\newcommand{\periodvalshort}{2.255354}
\newcommand{\rplanetval}{1.322\pm0.025}

\newcommand{\toplanetval}{2458149.40776\pm0.00091}

\newcommand{\teqplanetvalshort}{1561}
\newcommand{\mplanetval}{0.938^{+0.045}_{-0.042}}

\newcommand{\densplanetval}{0.504^{+0.038}_{-0.035}}

\newcommand{\loggplanetvalshort}{3.124}

\newcommand{\absolutervval}{-15.224\pm0.10}

\newcommand{\SPCteff}{5915\pm50}
\newcommand{\SPClogg}{4.45\pm0.10}
\newcommand{\SPCfeh}{-0.09\pm0.08}

\newcommand{\APFteff}{5775\pm110}
\newcommand{\APFfeh}{-0.03\pm0.06}

\begin{document}

\title{KELT-23A\MakeLowercase{b}: A Hot Jupiter Transiting a Near-Solar Twin Close to the {\it TESS} and {\it JWST} Continuous Viewing Zones}

\author{Daniel Johns}
\affiliation{Department of Physical Sciences, Kutztown University, Kutztown, PA 19530, USA}
\author{Phillip A.\ Reed}
\affiliation{Department of Physical Sciences, Kutztown University, Kutztown, PA 19530, USA}
\author{Joseph E.\ Rodriguez}
\affiliation{Center for Astrophysics \textbar \ Harvard \& Smithsonian, 60 Garden St, Cambridge, MA 02138, USA}
\author{Joshua Pepper}
\affiliation{Department of Physics, Lehigh University, 16 Memorial Drive East, Bethlehem, PA, 18015, USA}
\author{Keivan G.\ Stassun}
\affiliation{Department of Physics and Astronomy, Vanderbilt University, Nashville, TN 37235, USA}
\affiliation{Department of Physics, Fisk University, 1000 17th Avenue North, Nashville, TN 37208, USA}
\author{Kaloyan Penev}
\affiliation{Department of Physics, The University of Texas at Dallas, 800 West Campbell Road, Richardson, TX 75080-3021 USA}
\author{B.\ Scott Gaudi}
\affiliation{Department of Astronomy, The Ohio State University, 140 West 18th Avenue, Columbus, OH 43210, USA}
\author{Jonathan Labadie-Bartz}
\affiliation{Instituto de Astronomia, Geof\'isica e Ci\^encias Atmosf\'ericas, Universidade de S\~ao Paulo, Rua do Mat\~ao 1226, Cidade Universit\'aria, 05508-900, S\~ao Paulo, SP, Brazil}
\affiliation{Department of Physics \& Astronomy, University of Delaware, Newark, DE 19716, USA}
\author{Benjamin J.\ Fulton}
\affiliation{Caltech-IPAC/NExScI, Caltech, Pasadena, CA 91125 USA}
\author{Samuel N.\ Quinn}
\affiliation{Center for Astrophysics \textbar \ Harvard \& Smithsonian, 60 Garden St, Cambridge, MA 02138, USA}
\author{Jason D.\ Eastman}
\affiliation{Center for Astrophysics \textbar \ Harvard \& Smithsonian, 60 Garden St, Cambridge, MA 02138, USA}
\author{David R.\ Ciardi}
\affiliation{Caltech-IPAC/NExScI, Caltech, Pasadena, CA 91125 USA}
\author{Lea Hirsch}  
\affiliation{Kavli Institute for Particle Astrophysics and Cosmology, Stanford University, Stanford, CA 94305, USA}
\author{Daniel J.\ Stevens}
\affiliation{Department of Astronomy \& Astrophysics, The Pennsylvania State University, 525 Davey Lab, University Park, PA 16802, USA}
\affiliation{Center for Exoplanets and Habitable Worlds, The Pennsylvania State University, 525 Davey Lab, University Park, PA 16802, USA}
\author{Catherine P.\ Stevens}
\affiliation{Department of Physics, Westminster College, New Wilmington, PA 16172}
\author{Thomas E.\ Oberst}
\affiliation{Department of Physics, Westminster College, New Wilmington, PA 16172}
\author{David H.\ Cohen}
\affiliation{Department of Physics \& Astronomy, Swarthmore College, Swarthmore, PA 19081, USA}
\author{Eric L.\ N.\ Jensen}
\affiliation{Department of Physics \& Astronomy, Swarthmore College, Swarthmore, PA 19081, USA}
\author{Paul\ Benni}
\affiliation{Acton Sky Portal (private observatory), Acton, MA 01720,  USA}
\author{Steven Villanueva Jr.}
\affiliation{Department of Astronomy, The Ohio State University, 140 West 18th Avenue, Columbus, OH 43210, USA}
\author{Gabriel Murawski}
\affiliation{Gabriel Murawski Private Observatory}
\author{Allyson Bieryla}
\affiliation{Center for Astrophysics \textbar \ Harvard \& Smithsonian, 60 Garden St, Cambridge, MA 02138, USA}
\author{David W.\ Latham}
\affiliation{Center for Astrophysics \textbar \ Harvard \& Smithsonian, 60 Garden St, Cambridge, MA 02138, USA}
\author{Siegfried Vanaverbeke}
\affiliation{AstroLAB IRIS, Provinciaal Domein De Palingbeek, Verbrandemolenstraat 5, B-8902 Zillebeke, Ieper, Belgium}
\author{Franky Dubois}  
\affiliation{AstroLAB IRIS, Provinciaal Domein De Palingbeek, Verbrandemolenstraat 5, B-8902 Zillebeke, Ieper, Belgium}
\author{Steve Rau}  
\affiliation{AstroLAB IRIS, Provinciaal Domein De Palingbeek, Verbrandemolenstraat 5, B-8902 Zillebeke, Ieper, Belgium}
\author{Ludwig Logie}  
\affiliation{AstroLAB IRIS, Provinciaal Domein De Palingbeek, Verbrandemolenstraat 5, B-8902 Zillebeke, Ieper, Belgium}
\author{Ryan F.\ Rauenzahn}
\affiliation{Department of Physical Sciences, Kutztown University, Kutztown, PA 19530, USA}
\author{Robert A. Wittenmyer}
\affiliation{University of Southern Queensland, Centre for Astrophysics, West Street, Toowoomba, QLD 4350 Australia}

\author{Roberto Zambelli}
\affiliation{Societ\`{a} Astronomica Lunae, Italy}

\author{Daniel Bayliss}
\affiliation{Research School of Astronomy and Astrophysics, Mount Stromlo Observatory, Australian National University, Cotter Road, Weston, ACT, 2611, Australia}
\affiliation{Department of Physics, University of Warwick, Gibbet Hill Rd., Coventry, CV4 7AL, UK}
\author{Thomas G.\ Beatty}
\affiliation{Department of Astronomy and Steward Observatory, University of Arizona, Tucson, AZ 85721, USA}
\affiliation{Center for Exoplanets and Habitable Worlds, The Pennsylvania State University, 525 Davey Lab, University Park, PA 16802, USA}
\author{Karen A.\ Collins}
\affiliation{Center for Astrophysics \textbar \ Harvard \& Smithsonian, 60 Garden St, Cambridge, MA 02138, USA}
\author{Knicole D.\ Col\'on}
\affiliation{NASA Goddard Space Flight Center, Exoplanets and Stellar Astrophysics Laboratory (Code 667), Greenbelt, MD 20771, USA}
\author{Ivan A.\ Curtis}
\affiliation{Ivan Curtis Private Observatory}
\author{Phil Evans}
\affiliation{El Sauce Observatory, Chile}
\author{Joao Gregorio}
\affiliation{Atalaia Group \& CROW Observatory, Portalegre, Portugal}
\author{David James}
\affiliation{Event Horizon Telescope, Center for Astrophysics $|$ Harvard \& Smithsonian, MS-42, 60 Garden Street, Cambridge, MA 02138, USA}
\author{D.\ L.\ Depoy}  
\affiliation{George P.\ and Cynthia Woods Mitchell Institute for Fundamental Physics and Astronomy, Texas A\&M University, College Station, TX77843 USA}
\affiliation{Department of Physics and Astronomy, Texas A\&M university, College Station, TX 77843 USA}
\author{Marshall C.\ Johnson}
\affiliation{Department of Astronomy, The Ohio State University, 140 West 18th Avenue, Columbus, OH 43210, USA}
\author{Michael D.\ Joner}
\affiliation{Department of Physics and Astronomy, Brigham Young University, Provo, UT 84602, USA}
\author{David H.\ Kasper}
\affiliation{Department of Physics \& Astronomy, University of Wyoming, 1000 E University Ave, Dept 3905, Laramie, WY 82071, USA}
\author{Somayeh Khakpash}
\affiliation{Department of Physics, Lehigh University, 16 Memorial Drive East, Bethlehem, PA, 18015, USA}
\author{John F.\ Kielkopf}
\affiliation{Department of Physics and Astronomy, University of Louisville, Louisville, KY 40292 USA}
\author{Rudolf B.\ Kuhn}
\affiliation{South African Astronomical Observatory, PO Box 9, Observatory, 7935, Cape Town, South Africa}
\affiliation{Southern African Large Telescope, PO Box 9, Observatory, 7935, Cape Town, South Africa}
\author{Michael B.\ Lund}
\affiliation{Caltech-IPAC/NExScI, Caltech, Pasadena, CA 91125 USA}
\affiliation{Department of Physics and Astronomy, Vanderbilt University, Nashville, TN 37235, USA}
\author{Mark Manner}
\affiliation{Spot Observatory, Nashville, TN 37206, USA}
\author{Jennifer L.\ Marshall}  
\affiliation{George P.\ and Cynthia Woods Mitchell Institute for Fundamental Physics and Astronomy, Texas A\&M University, College Station, TX77843 USA}
\affiliation{Department of Physics and Astronomy, Texas A\&M university, College Station, TX 77843 USA}
\author{Kim K.\ McLeod}
\affiliation{Department of Astronomy, Wellesley College, Wellesley, MA 02481, USA}
\author{Matthew T.\ Penny}
\affiliation{Department of Astronomy, The Ohio State University, 140 West 18th Avenue, Columbus, OH 43210, USA}
\author{Howard Relles}
\affiliation{Center for Astrophysics \textbar \ Harvard \& Smithsonian, 60 Garden St, Cambridge, MA 02138, USA}
\author{Robert J.\ Siverd}
\affiliation{Department of Physics and Astronomy, Vanderbilt University, Nashville, TN 37235, USA}
\author{Denise C.\ Stephens}
\affiliation{Department of Physics and Astronomy, Brigham Young University, Provo, UT 84602, USA}
\author{Chris Stockdale}
\affiliation{Hazelwood Observatory, Churchill, Victoria, Australia}
\author{Thiam-Guan Tan}
\affiliation{Perth Exoplanet Survey Telescope}
\author{Mark Trueblood}  
\affiliation{Winer Observatory, PO Box 797, Sonoita, AZ 85637, USA}
\author{Pat Trueblood}  
\affiliation{Winer Observatory, PO Box 797, Sonoita, AZ 85637, USA}
\author{Xinyu Yao}
\affiliation{Department of Physics, Lehigh University, 16 Memorial Drive East, Bethlehem, PA, 18015, USA}

\shorttitle{KELT-23A\MakeLowercase{b}}
\shortauthors{Johns et al.}

\begin{abstract}
	We announce the discovery of KELT-23Ab, a hot Jupiter transiting the relatively bright ($V=10.3$) star BD+66 911 (TYC 4187-996-1), and characterize the system using follow-up photometry and spectroscopy. A global fit to the system yields host-star properties of $\teff=\teffstarval$~K, $M_*=\mstarval \msun$, $R_*=\rstarval \rsun$, $L_*=\lstarval \lsun$, $\loggstar=\loggstarval$ (cgs), and $\feh=\fehstarval$.  KELT-23Ab is a hot Jupiter with mass $M_P=\mplanetval \mj$, radius $R_P=\rplanetval \rj$, and density $\rho_P=\densplanetval$ g cm$^{-3}$. Intense insolation flux from the star has likely caused KELT-23Ab to become inflated. The time of inferior conjunction is $T_0=\toplanetval~\bjdtdb$ and the orbital period is $P=\periodval$ days. There is strong evidence that KELT-23A is a member of a long-period binary star system with a less luminous companion, and due to tidal interactions, the planet is likely to spiral into its host within roughly a Gyr. This system has one of the highest positive ecliptic latitudes of all transiting planet hosts known to date, placing it near the {\it Transiting Planet Survey Satellite} and {\it James Webb Space Telescope} continuous viewing zones. Thus we expect it to be an excellent candidate for long-term monitoring and follow-up with these facilities.
\end{abstract}





\section{Introduction}
The first known transiting exoplanet, HD 209458b \citep{Charbonneau:2000, Henry:2000}, is now but one of over 3000 confirmed exoplanets. Most of these planets were discovered by the {\it Kepler} mission \citep{Borucki:2010, Howell:2014, Coughlin:2016}, and the first of many new discoveries are now coming from the \textit{Transiting Exoplanet Survey Satellite} (TESS; \citealt{Ricker:2015, Huang:2018, Vanderspek:2019}). The short period, massive planet population known as hot Jupiters produce signals that are large enough to be detected from ground-based transit surveys such as the Trans-atlantic Exoplanet Survey (TrES; \citealt{Alonso:2004}), the XO project \citep{McCullough:2005}, the Wide Angle Search for Planets (WASP; \citealt{Pollacco:2006}), the Hungarian Automated Telescope Network (HATNet; \citealt{Bakos:2004}), the Qatar Exoplanet Survey (QES; \citealt{Alsubai:2013}), the Next-Generation Transit Survey (NGTS; \citealt{Wheatley:2013}), the Multi-site All Sky Camera (MASCARA; \citealt{Talens:2017a}), and the Kilodegree Extremely Little Telescope (KELT; \citealt{Pepper:2003, Pepper:2007, Pepper:2012}).

The majority of the known Kepler planets orbit faint stars, but have relatively long periods compared to the transiting planets found by ground-based surveys due to the long duration and continuous observations of Kepler's prime survey. Indeed, ground-based surveys are typically only sensitive to planets with periods of less than roughly ten days due primarily to the effects of weather, the diurnal cycle, and the atmosphere, resulting in a non-continuous cadence and poor sensitivity to longer planets, but find planets around stars that are much brighter than Kepler due to their smaller apertures and ability to monitor much larger regions of the sky. One such survey, KELT\footnote{\href{https://keltsurvey.org}{https://keltsurvey.org}}, is a pair of telescopes situated in the Northern and Southern Hemispheres, in Arizona (KELT-North) and South Africa (KELT-South) respectively. KELT-North and South each have apertures 42mm in diameter with a $26\arcdeg\times26\arcdeg$ field of view and a $23\arcsec$ pixel scale. KELT-North and KELT-South began collecting data in 2005 and 2009, respectively. Between them, they survey roughly 61\% of the sky at 10-20 minute cadence with about 1\% photometric precision down to V$\sim$11. KELT is most sensitive to giant planets transiting stars between 8th and 11th magnitude in the $V$-band, a range that is often missed by radial velocity surveys and many other transit surveys, because either the hosts are too faint and numerous (for RV), or because stars at the brighter end of this magnitude range are saturated for ground-based surveys with larger apertures than KELT. This range implies KELT is biased toward detecting inflated, giant planets \citep{Gaudi:2005} around hot, luminous stars \citep{Bieryla:2015}. Since the targets selected by KELT are relatively bright they present a valuable opportunity for atmospheric characterization. KELT has discovered over 20 exoplanets, the latest being KELT-21b \citep{Johnson:2017} from KELT-North and KELT-22Ab \citep{Bartz:2018} from KELT-South.

The inflated giant planets sought by KELT are hot Jupiters. They typically have masses greater than 0.25 $M_{\textrm J}$, radii between 1 and 2 $R_{\textrm J}$, and periods less than 10 days. They present a unique opportunity to study the evolution and formation of exoplanetary systems. Hot Jupiters are fairly easy to detect, however \cite{Wright:2012} and \cite{Fressin:2013} estimate that only around 1\% of Sun-like stars host hot Jupiters, making them relatively rare compared to smaller planets. The frequency and depth of photometric transits of the hot Jupiters make them optimal for detection by the above-mentioned surveys. KELT plays an important role discovering and confirming some of the most promising hot Jupiters for follow-up observations by the {\it Hubble Space Telescope} (HST), {\it Spitzer Space Telescope} \citep[see][]{Beatty:2018}, and the {\it James Webb Space Telescope} (JWST).

Gas giants also present a unique opportunity to study atmospheric characteristics in planets outside of our solar system. Transmission spectroscopy is the illumination through the planet's atmosphere by the host star during a transit. This effect allows us to study the atmospheric composition, wind speeds, haze, rotation rate, etc. \citep{Charbonneau:2002, Brogi:2016, Sing:2016}. In addition to transmission spectroscopy, secondary eclipses give insight into the planet's albedo and thermal emission in the near infrared \citep{Garhart:2019}. 

Here we report the discovery of KELT-23Ab, a hot Jupiter orbiting a Sunlike star. KELT-23A is a bright ($V=10.25$), near-Solar metallicity ($\feh=\fehstarvalshort$), main-sequence, G2-type host whose composition may give insight into planetary formation processes around Sunlike stars. \cite{Melendez:2009} found that the Sun's composition has a distinguishing abundance pattern that is not shared by solar analogs known to host hot Jupiters.  More recently, \cite{Bedell:2018} analyzed the abundance ratios of Sunlike exoplanet hosts within the Solar neighborhood and found that a majority of them have similar abundance patterns. A differential abundance analysis may give clues as to why the Solar system does not host a hot Jupiter, yet KELT-23A does. Our analysis suggests that KELT-23Ab is on a decaying orbit and is spiraling in toward the host star, a circumstance that could provide an opportunity to better understand the late stages of giant planet orbital evolution around a Sunlike star.

In Section \ref{sec:Phot_Spec_AO} we describe the initial discovery light curve, follow-up photometry and spectroscopy, and adaptive optics observations. In Section \ref{sec:star_props} we use follow-up observations to characterize the host star. In Section \ref{sec:analysis} we fit a global model to the planetary system to measure the planet's parameters. We also search for transit timing variations and consider the likelihood of a wide binary companion to KELT-23A.  We conclude that the planetary host is the brighter of the two stars and have therefore given it the appropriate designation (``A'') in its name. Finally, in Sections \ref{sec:false_pos} and \ref{sec:discussion} we rule out the possibility of a false positive scenario and discuss the tidal evolution of the system, stellar abundances, and future observations.

\section{Discovery and Follow-Up Observations} \label{sec:Phot_Spec_AO}

\subsection{KELT-North Observations and Photometry}\label{sec:keltobs}
\begin{figure}[ht!]
    \vspace{.0in}
    \centering\includegraphics[width=1.0\columnwidth, trim = 0 2.9in 0 0]{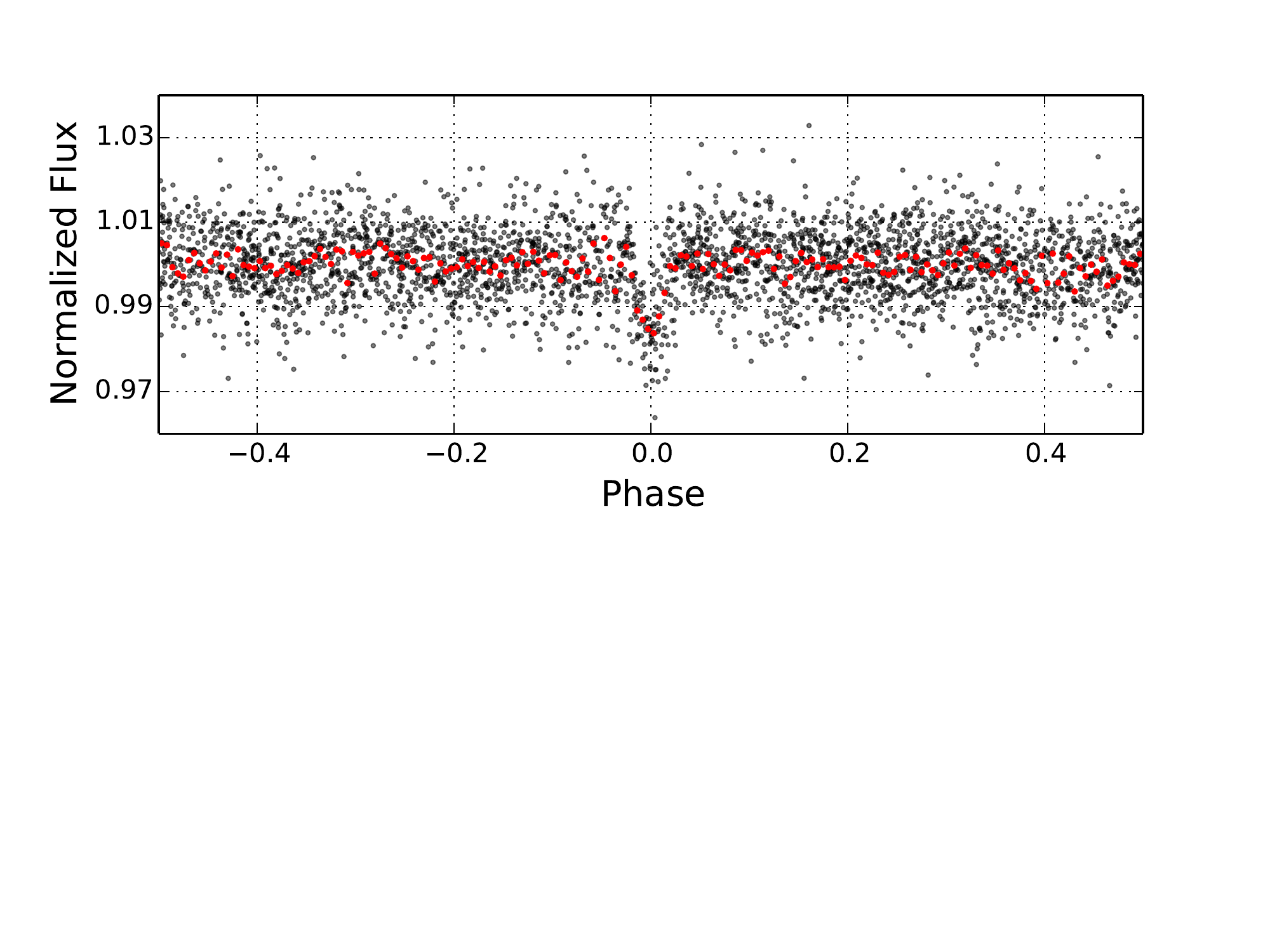}
    \caption{\footnotesize Discovery light curve of KELT-23A containing 3308 observations from the KELT-North telescope phase-folded on the discovery period of 2.2552347 days. The red points are the data binned on an 18-minute time scale. }
    \label{fig:DiscoveryLC}
\end{figure}

KELT-23A is located in KELT-North survey field 22, a $26\arcdeg\times26\arcdeg$ region centered at ($\alpha_{\rm J2000}=16^h03^m12\fs0$,~$\delta_{\rm J2000}=+57\arcdeg00\arcmin00\farcs0$). We observed the field a total of 3308 times from 2012 February to 2017 August. TYC 4187-996-1 was identified as a transit candidate within the field at $\alpha=15^h28^m35\fs1926,~\delta=+66\arcdeg21\arcmin31\farcs544$. The initial discovery light curve, shown in Figure \ref{fig:DiscoveryLC}, contains a significant Box-Least-Squares \citep{Kovacs:2002} signal with a period of 2.2552375 days, a transit duration of 1.80 hours, and a transit depth of 12.8 mmag. The KELT image analysis and light curve production pipeline is described in detail in \citet{Siverd:2012}. Table \ref{tbl:LitProps} shows photometric and astrometric properties of KELT-23A.

\begin{table}
\footnotesize
\centering
\caption{Literature Properties for KELT-23A}
\begin{tabular}{llcc}
  \hline
  \hline
Other Names\dotfill & 
       \\
      & \multicolumn{3}{l}{BD+66 911}				\\
	  & \multicolumn{3}{l}{TYC 4187-996-1}\\
	  & \multicolumn{3}{l}{TIC 458478250}\\
\\
\hline
Parameter & Description & Value & Ref. \\
\hline
$\alpha_{\rm J2000}$\dotfill	&Right Ascension (RA)\dotfill & $15^h28^m35\fs1926$			& 1	\\
$\delta_{\rm J2000}$\dotfill	&Declination (Dec)\dotfill & $+66\arcdeg21\arcmin31\farcs544$			& 1	\\
$\beta_{\rm J2000}$\dotfill & Ecliptic Latitude\dotfill & +75\arcdeg6\arcmin43\arcsec.21 & 1\\
$\lambda_{\rm J2000}$\dotfill & Ecliptic Longitude\dotfill & 163\arcdeg17\arcmin10\arcsec.38 & 1\\
\\
$B_{\rm T}$\dotfill			&Tycho $B_{\rm T}$ mag.\dotfill & $11.029 \pm 0.049$		& 2	\\
$V_{\rm T}$\dotfill			&Tycho $V_{\rm T}$ mag.\dotfill & $10.376 \pm 0.039$		& 2	\\

$B$\dotfill		& APASS Johnson $B$ mag.\dotfill	& $10.973 \pm 0.021$		& 3	\\
$V$\dotfill		& APASS Johnson $V$ mag.\dotfill	& $10.253 \pm 0.024$	& 3	\\

$g'$\dotfill		& APASS Sloan $g'$ mag.\dotfill	&  $10.647 \pm 0.026$		& 3	\\
$r'$\dotfill		& APASS Sloan $r'$ mag.\dotfill	&  $10.114 \pm 0.080$      & 3	\\
$i'$\dotfill		& APASS Sloan $i'$ mag.\dotfill	&  $9.980 \pm 0.070$	& 3	\\
\\
$J$\dotfill			& 2MASS $J$ mag.\dotfill & $9.208 \pm 0.032$		& 4	\\
$H$\dotfill			& 2MASS $H$ mag.\dotfill & $\geq8.951^*$	& 4	\\
$K$\dotfill			& 2MASS $K$ mag.\dotfill & $\geq8.904^*$	& 4	\\
\\
\textit{WISE1}\dotfill		& \textit{WISE1} mag.\dotfill & $8.746 \pm 0.022$		& 5	\\
\textit{WISE2}\dotfill		& \textit{WISE2} mag.\dotfill & $8.778 \pm 0.020$		& 5 \\
\textit{WISE3}\dotfill		& \textit{WISE3} mag.\dotfill & $8.778 \pm 0.019$		& 5	\\
\textit{WISE4}\dotfill		& \textit{WISE4} mag.\dotfill & $8.672 \pm 0.189$		& 5	\\
\\
$\mu_{\alpha}$\dotfill		& Gaia DR2 proper motion\dotfill & 0.434 $\pm$ 0.039 		& 1 \\
                    & \hspace{3pt} in RA (mas yr$^{-1}$)	& & \\
$\mu_{\delta}$\dotfill		& Gaia DR2 proper motion\dotfill 	&  -12.217 $\pm$ 0.041 &  1 \\
                    & \hspace{3pt} in DEC (mas yr$^{-1}$) & & \\
$\Pi$\dotfill & Gaia DR2 Parallax (mas) \dotfill &$7.973\pm0.021$& 1\dag \\ 
\\
$RV$\dotfill & Systemic radial \hspace{9pt}\dotfill  & $\absolutervval$ & \S\ref{sec:spectroscopy} \\
     & \hspace{3pt} velocity (\kms)  & & \\
$v\sin{i_\star}$\dotfill &  Stellar rotational \hspace{7pt}\dotfill &  $\vsinistarval$ & \S\ref{sec:spec_analysis} \\
& \hspace{3pt} velocity (\kms)  & & \\
Sp. Type\dotfill & Spectral Type\dotfill & G2V & \S\ref{sec:star_props} \\
Age\dotfill & Age (Gyr)\dotfill & $\agestarval$ & \S\ref{sec:st_age} \\
$d_{\star}$\dotfill& Distance (pc) \dotfill & $\diststarval$ & 1\dag \\
$A_V$\dotfill & Visual extinction (mag)\dotfill & $0.075\pm0.015$ & \S\ref{sec:SED} \\
$U^{\dag\dag}$\dotfill   & Space motion (\kms)\dotfill & $17.32 \pm 0.03$ & \S\ref{sec:UVW} \\
$V$\dotfill       & Space motion (\kms)\dotfill & $0.95 \pm 0.07$ & \S\ref{sec:UVW} \\
$W$\dotfill       & Space motion (\kms)\dotfill & $-0.90 \pm 0.07$ & \S\ref{sec:UVW} \\
\hline
\hline
\end{tabular}
\begin{flushleft} 
 \footnotesize{ \textbf{\textsc{NOTES:}} All photometric apertures are blended with the partner within 4$\farcs$5. References are: $^1$\citet{Brown:2018} Gaia DR2 http://gea.esac.esa.int/archive/, $^2$\citet{Hog:2000},$^3$\citet{Henden:2015}, $^4$\citet{Cutri:2003}, $^5$\citet{Cutri:2012}. $^*$ 2MASS magnitude contains the "U" quality flag. \dag Gaia DR2 parallax and distance after correcting for the systematic offset of $-0.082$~mas as described in \citet{Stassun:2018}. \dag\dag~$U$ is taken to be positive in the direction of the galactic center.
}
\end{flushleft}
\label{tbl:LitProps}
\end{table}

\subsection{Photometric Time-series Follow-up} \label{sec:phot}
\begin{figure}[!t]
\vspace{0.3in}
\centering\includegraphics[width=0.99\linewidth, trim = 1.5in 0in 0.5in 0.1in, clip]{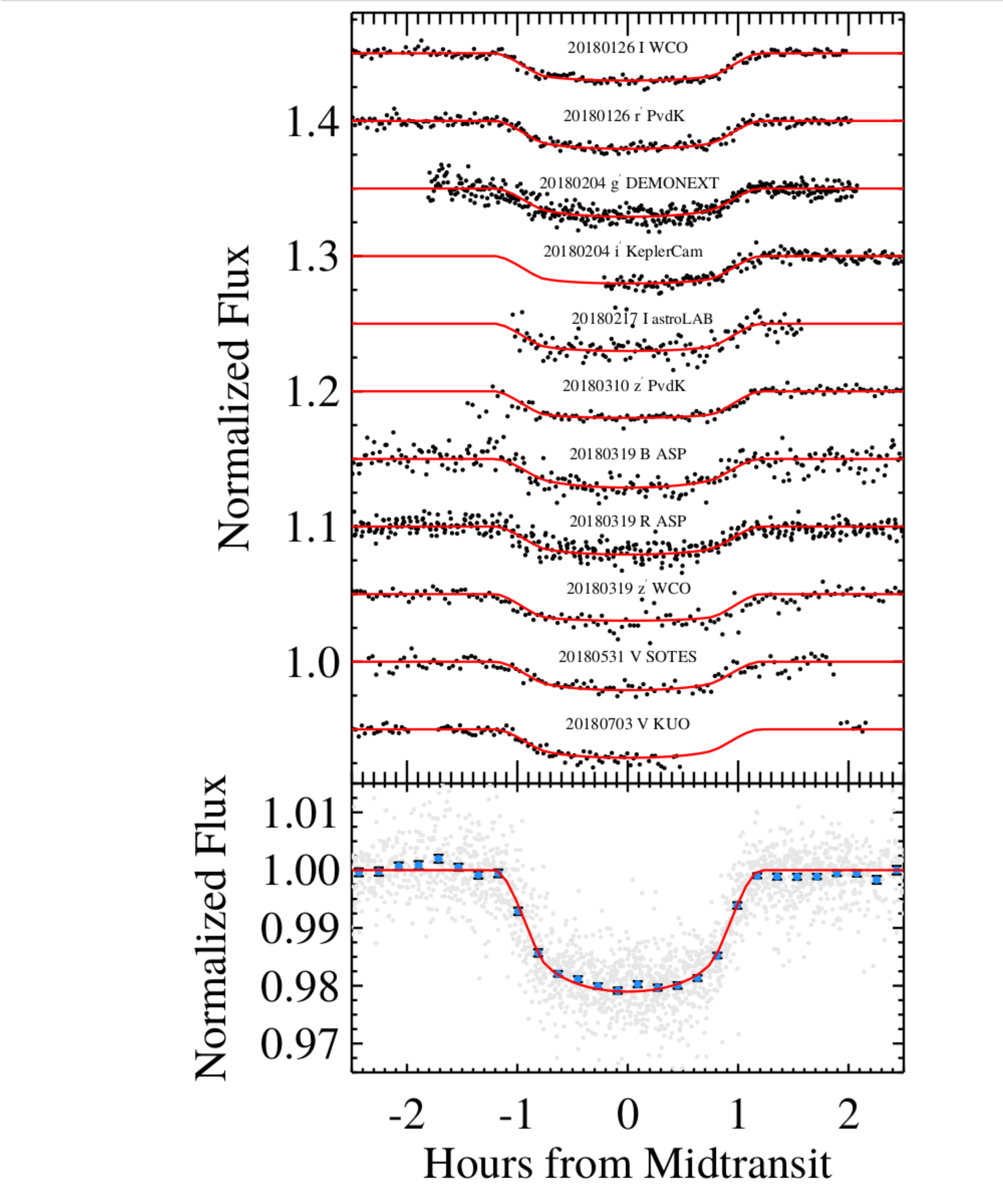}  
\caption{ {\it Top}: Transit light curves of KELT-23Ab from the KELT-North FUN. Observations are offset by an arbitrary amount and are plotted in chronological order with the observatory's acronym (see Table \ref{tab:Follow-up}) and the date of observation above the light curve. Black dots represent observed relative flux. The red lines represent the fit to each data set. {\it Bottom}: All follow-up light curves combined and binned (blue squares). The red line represents the fit to the data. This combined light curve is strictly used to highlight the statistical significance of follow-up observations and is not used for analysis.}
\label{fig:All_LCs}
\end{figure}

The KELT-Follow-Up Network (FUN) is a collaboration of around 60 observatories spread throughout the Northern and Southern hemispheres.\footnote{A more complete listing of KELT-FUN observatories can be found at \href{https://keltsurvey.org/}{https://keltsurvey.org/}}~
The KELT-FUN follow-up photometry is used to better constrain the transit parameters, eliminate false-positives, and determine accurate ephemerides. KELT-FUN members plan photometric follow-up observations through the use of TAPIR \citep{Jensen:2013}, a web-based transit prediction calculator. KELT-FUN members generally use the AstroImageJ (AIJ) package \citep{Collins:2013, Collins:2017}\footnote{\href{https://www.astro.louisville.edu/software/astroimagej/}{https://www.astro.louisville.edu/software/astroimagej/}} to reduce and perform a preliminary analysis of their light curves. We obtained 11 follow-up transits between 26 January 2018 and 03 July 2018 from the observatories listed in Table \ref{tab:Follow-up}. The specifications listed in Table \ref{tab:Follow-up} are taken from \cite{Collins2018}, which is the most current listing of KELT-FUN instrument specifications. The follow-up light curves span the Johnson $BVRI$ and Sloan $g'r'i'z'$ filter sets. As part of the follow-up confirmation process the transits in different bandpasses were checked for chromatic variations. No chromatic variations were found. The individual light curves are shown in Figure \ref{fig:All_LCs}. 

The KELT-FUN light curves were detrended by various parameters such as airmass, full-width-half-max (FWHM), X position, Y position, sky brightness, and total comparison star counts. To determine the best set of detrending parameters we used the Bayesian Information Criterion (BIC) of a preliminary fit to the detrended data. Detrending parameters that produced a drop in BIC of 10 or greater were selected for the global fit (Section \ref{sec:globalfit}). These free parameters can be found in Table \ref{tab:Follow-up}. In all cases, airmass was selected as a detrending parameter due to the pervasive airmass-dependent systematic effects seen in differential photometry. Other parameters such as X and Y target pixel position were evaluated as they can induce trends in the light curve due to flatfielding. Observatories that perform a meridian flip during observations additionally have the option of detrending by meridian flip in AIJ as a meridian flip can cause a shift in position of the target. Meridian flip detrending was not found to be necessary for any KELT-FUN light curves in the current analysis.

\begin{table*}
 \footnotesize
 \centering
 \setlength\tabcolsep{1.5pt}
 \caption{Photometric follow-up observations of KELT-23A\MakeLowercase{b}}
 \begin{tabular}{lccccccccc}
   \hline
   \hline
Observatory & Location & Aperture~~ & Plate scale& Date & Filter~~ & Exposure~~ & RMS$^a$ & Detrending Parameters$^b$ & Transit \\ 
& & (m) & ($\rm \arcsec~pix^{-1}$)& (UT 2018) & & Time (s) & (10$^{-3}$) & & Coverage \\
\hline
WCO & PA, USA & 0.35 & 0.45 & 26 January & $I$ & 90 & 2.42 & Airmass, Total Counts, Y & Full Transit \\
PvdK & PA, USA & 0.61 & 0.38 & 26 January & $r'$ & 60 & 2.67 & Airmass, FWHM & Full Transit \\
DEMONEXT & AZ, USA & 0.50 & 0.90 & 03 February & $g'$ & 30 & 4.95 & Airmass, X & Full Transit \\
KeplerCam & AZ, USA & 1.20 & 0.37 & 04 February & $i'$ & 15 & 3.58 & Airmass & Mid-Transit, Egress \\
astroLAB & Zillebeke, Belgium & 0.684 & 1.86 & 17 February & $I$ & 5 & 6.01 &  Airmass, Sky Brightness & Partial Ingress, Egress \\
PvdK & PA, USA & 0.61 & 0.38 & 10 March & $z'$ & 60 & 2.86 & Airmass, Y & Full Transit \\
ASP & MA, USA & 0.355 & 0.69 & 19 March & $B$ & 30 & 5.80 & Airmass, Total Counts, X & Full Transit \\
ASP & MA, USA & 0.355 & 0.69 & 19 March & $R$ & 10 & 4.78 & Airmass, Total Counts & Full Transit \\
WCO & PA, USA & 0.35 & 0.45 & 19 March & $z'$ & 120 & 4.60 & Airmass & Full Transit \\
SOTES & Suwalki, Poland & 0.10 & 1.65 & 31 May & $V$ & 120 & 4.05 & Airmass, Total Counts & Full Transit \\
KUO & PA, USA & 0.61 & 0.72 & 03 July & $V$ & 90 & 2.55 & Airmass, Total Counts  & Ingress, Mid-Transit, Post-Egress \\
\hline
\hline

\end{tabular}
\begin{flushleft}
  \footnotesize{\textbf{\textsc{NOTES:}}
  $^a$The root-mean-square scatter of the best-fit transit model residuals. 
  $^b$Photometric parameters allowed to vary in global fit and described in the text.
  
  Abbreviations: WCO: Westminster College Observatory; PvdK: Peter van de Kamp Observatory; DEMONEXT: DEdicated MONitor of EXotransits and Transients; AstroLAB: public observatory astroLAB IRIS; ASP: Acton Sky Portal; SOTES: Gabriel Murawski Private Observatory; KUO: Kutztown University Observatory.}
\end{flushleft}
\label{tab:Follow-up}
\end{table*}

\subsection{Spectroscopic Follow-up} \label{sec:spectroscopy}

\subsubsection{TRES at FLWO}

We obtained 21 spectra from the Tillinghast Reflector \'Echelle Spectrograph (TRES\footnote{\href{http://tdc-www.harvard.edu/instruments/TRES}{http://tdc-www.harvard.edu/instruments/TRES}}; \citealt{Szentgyorgyi:2007, Furesz:2008}) on the 1.5 m telescope at the Fred Lawrence Whipple Observatory on Mount Hopkins, Arizona, USA. We observed KELT-23A with TRES between UT 18 February 2018 and UT 07 June 2018. TRES is a $2.3\arcsec$ (as projected on the sky) fiber-fed \'echelle spectrograph with a wavelength dispersion between $3900-9100$\AA~and a resolving power of R$\sim44,000$. Radial velocities were obtained through the procedure described below. These spectra were also used to calculate Bisector Spans (BSs) for use in false-positive analysis in Section \ref{sec:false_pos}. The procedure for calculating BSs can be found in detail in \cite{Buchhave:2010}. The TRES RVs and BSs can be seen in Table \ref{tab:Spectra}.

\begin{figure}
	\vspace{.0in}
	\includegraphics[width=1\linewidth, page=2, trim={2.5cm 13cm 8.5cm 8cm}, clip]{KELT-23_rv.pdf}
	\includegraphics[width=1\linewidth, page=1, trim={2.5cm 13cm 8.5cm 8cm}, clip]{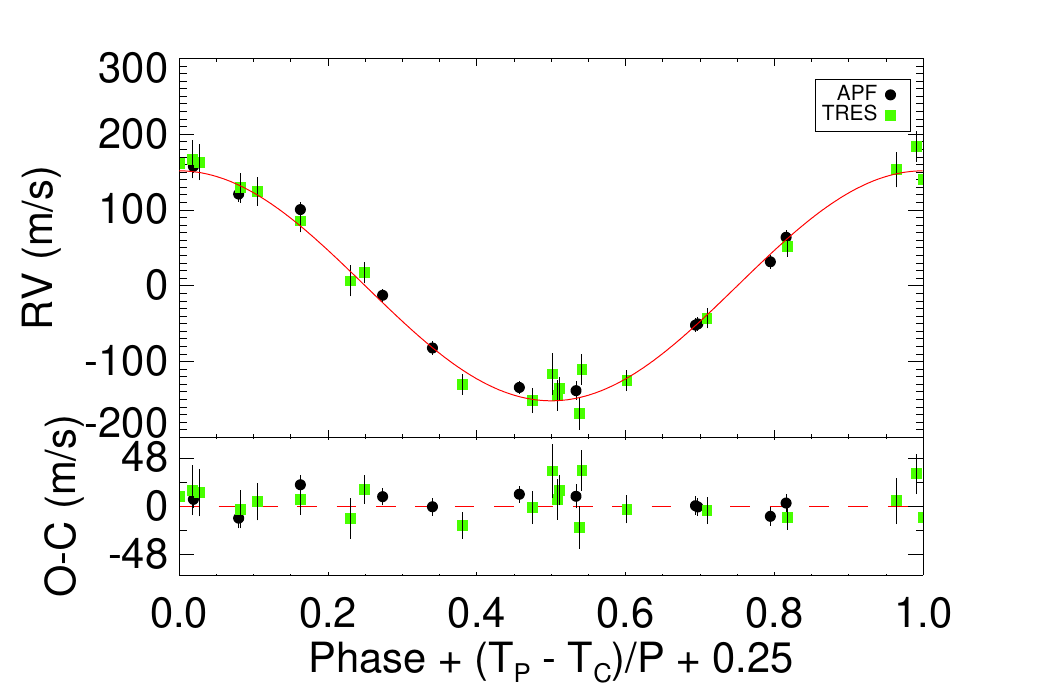}
	\caption{{\it Top}: TRES and APF radial velocity measurements. {\it Bottom}: TRES and APF radial velocity measurements phase folded to the period determined by the global fit ($P=2.25477$ days). Black points represent the APF radial velocities and their uncertainties. Green dots represent the TRES radial velocities and their uncertainties. The red line represent the global fit to the system determined in Section \ref{sec:globalfit}. The bottom panels show the residuals of the global fit.}
	\label{fig:TRES_RVs} 
\end{figure}

Relative RVs from TRES are found by first designating the spectrum with the highest Signal-to-Noise ratio (SNR) as the reference spectrum. Each individual spectrum is then cross-correlated with the reference. This cross-correlation utilizes the portion of the spectrum between $4300-5660$\AA. To obtain absolute RVs from TRES the reference spectrum's Mg b region (5190\AA) is cross correlated with the appropriate \cite{Kurucz:1992} stellar atmosphere spectrum. From there, absolute RVs are adjusted to the International Astronomical Union (IAU) Radial Velocity Standard Star system \citep{Stefanik:1999}. This correction is of the order $-0.61$ km s$^{-1}$ and serves to correct for the non-inclusion of gravitational redshift in the model spectrum. The absolute RV of the KELT-23A system was found to be $\absolutervval$ km s$^{-1}$, where the uncertainty is a result of the residual systematic error in the transformation to the IAU system. The TRES RVs are plotted in Figure \ref{fig:TRES_RVs} in green.

\subsubsection{APF at Lick Observatory}
We obtained 11 spectra of KELT-23A between UT 17 February 2018 and UT 20 June 2018 from the Levy spectrograph on the 2.4-meter Automated Planet Finder (APF\footnote{\href{https://www.ucolick.org/public/telescopes/apf.html}{https://www.ucolick.org/public/telescopes/apf.html}}: \citealt{Vogt:2014}) telescope at Lick Observatory, Mount Hamilton, CA. The APF spectra have a resolution of R$\sim100,000$ and are obtained through a $1\arcsec\times3\arcsec$ slit. Observations were done through an iodine cell, which imprints a forest of absorption lines onto the stellar spectrum to serve as a wavelength and instrumental broadening reference. The procedure for obtaining radial velocities from APF spectra differs from the procedure for TRES and is described in great detail in \cite{Fulton:2015}. The method of extracting RVs from the APF spectra involves deconvolving the stellar spectrum from the iodine cell's absorption line forest. The deconvolved stellar spectrum is then used to evaluate the doppler shift and model the instrumental point-spread function \citep{Butler:1996}.

\begin{figure}
    \centering
    \includegraphics[width=1\linewidth]{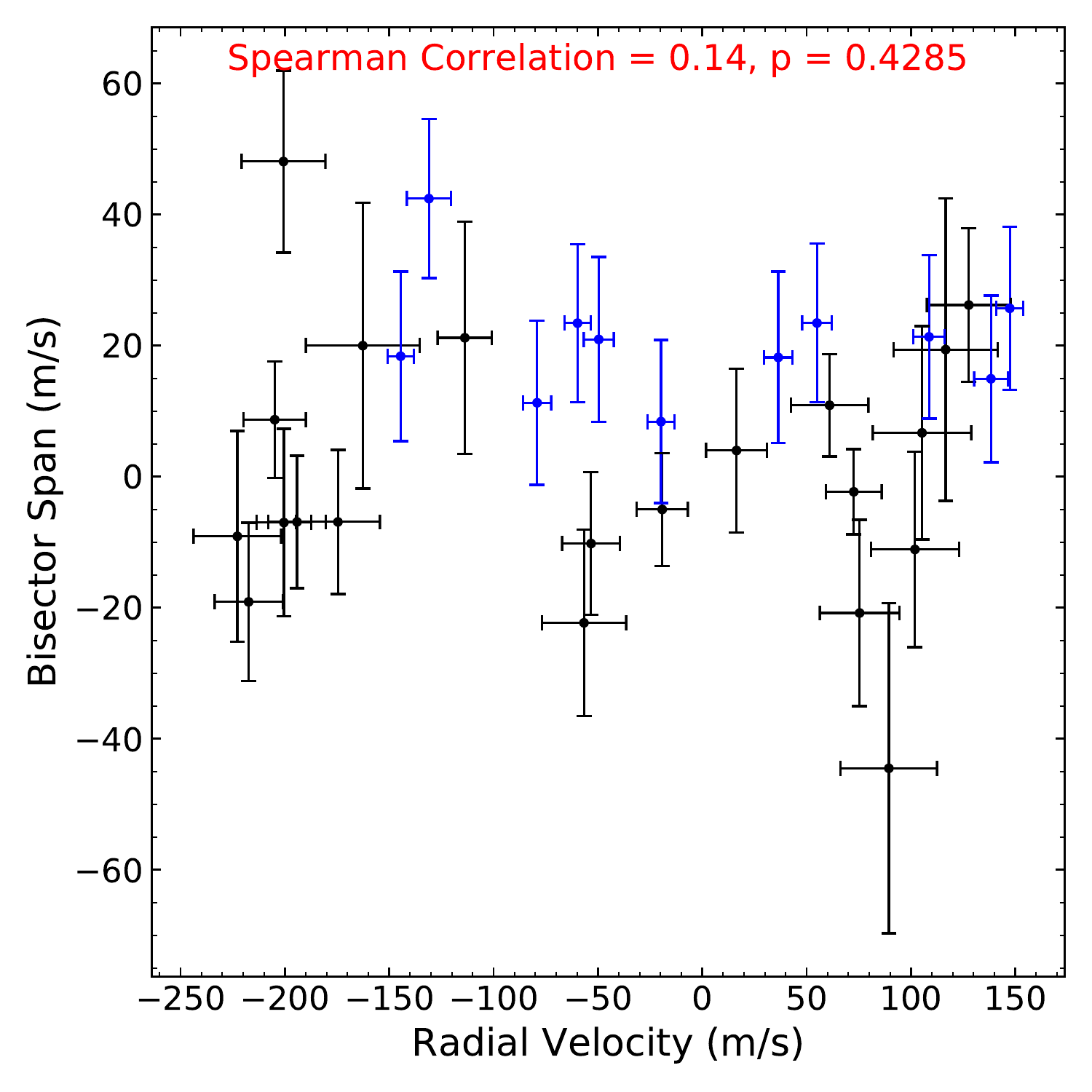}
    \caption{Bisector Spans for the TRES (black) and APF (blue) RV spectra plotted against the RV values. There is no statistically significant correlation between these quantities.}
    \label{fig:bisector_spans}
\end{figure}

\subsection{High-Contrast Imaging}\label{sec:ao}
As part of our standard process for validating transiting exoplanets, we observed KELT-23A with infrared high-resolution adaptive optics (AO) imaging at Keck Observatory. The high-resolution AO imaging has demonstrated the ability to resolve faint stellar companions, as in the case of the hierarchical triple KELT-21 system \citep{Johnson:2018}. The Keck Observatory observations were made with the NIRC2 instrument on Keck-II behind the natural guide star AO system. The observations were made on 2018~Jun~06 in the standard 3-point dither pattern that is used with NIRC2 to avoid the left lower quadrant of the detector which is typically noisier than the other three quadrants. The dither pattern step size was $3\arcsec$ and was repeated twice, with each dither offset from the previous dither by $0.5\arcsec$.  

\begin{figure}
    \centering
    \includegraphics[width=\linewidth]{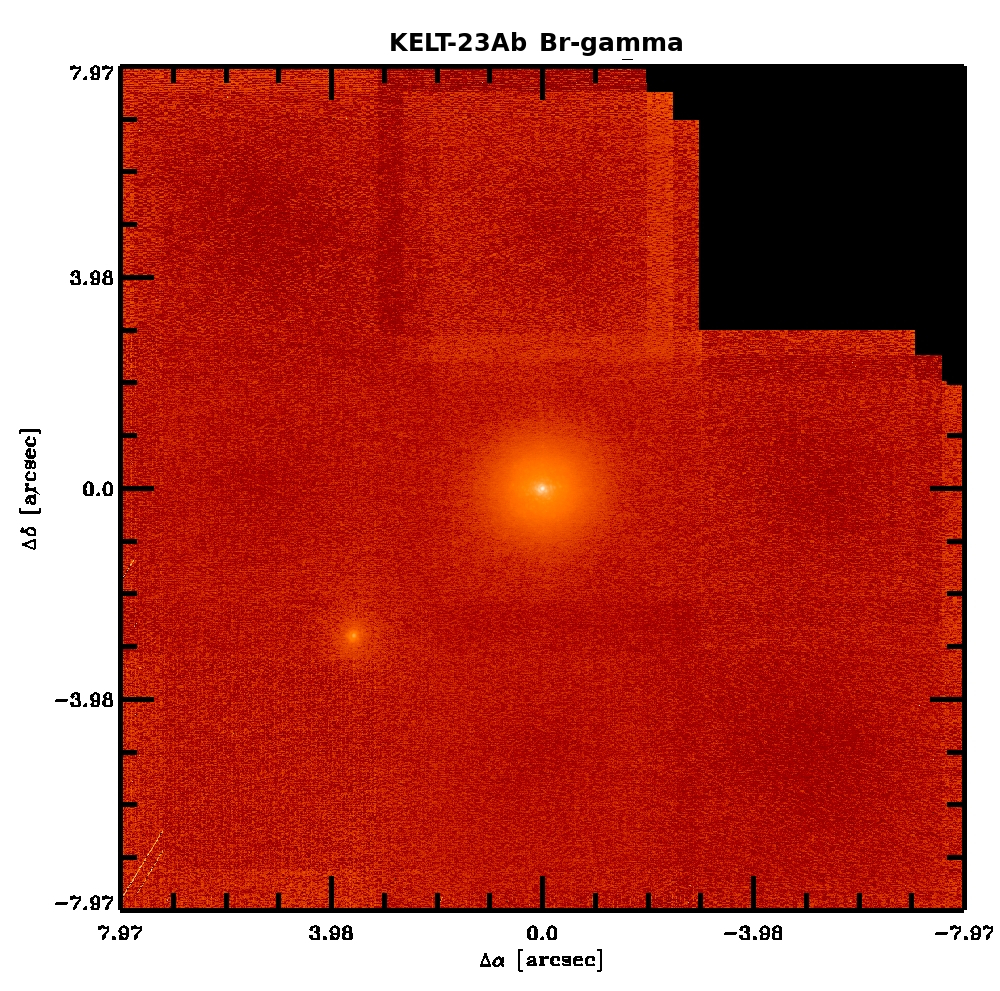}
    \caption{Full field of view image of the final combined dither pattern for the Keck adaptive optics imaging. The image clearly shows both known 2MASS stars but no other stars within $\sim10$\arcsec.}
    \label{fig:AO_image}
\end{figure}

\begin{figure}
    \centering
    \includegraphics[width=1\linewidth, trim={8cm 2cm 2cm 4cm}, clip]{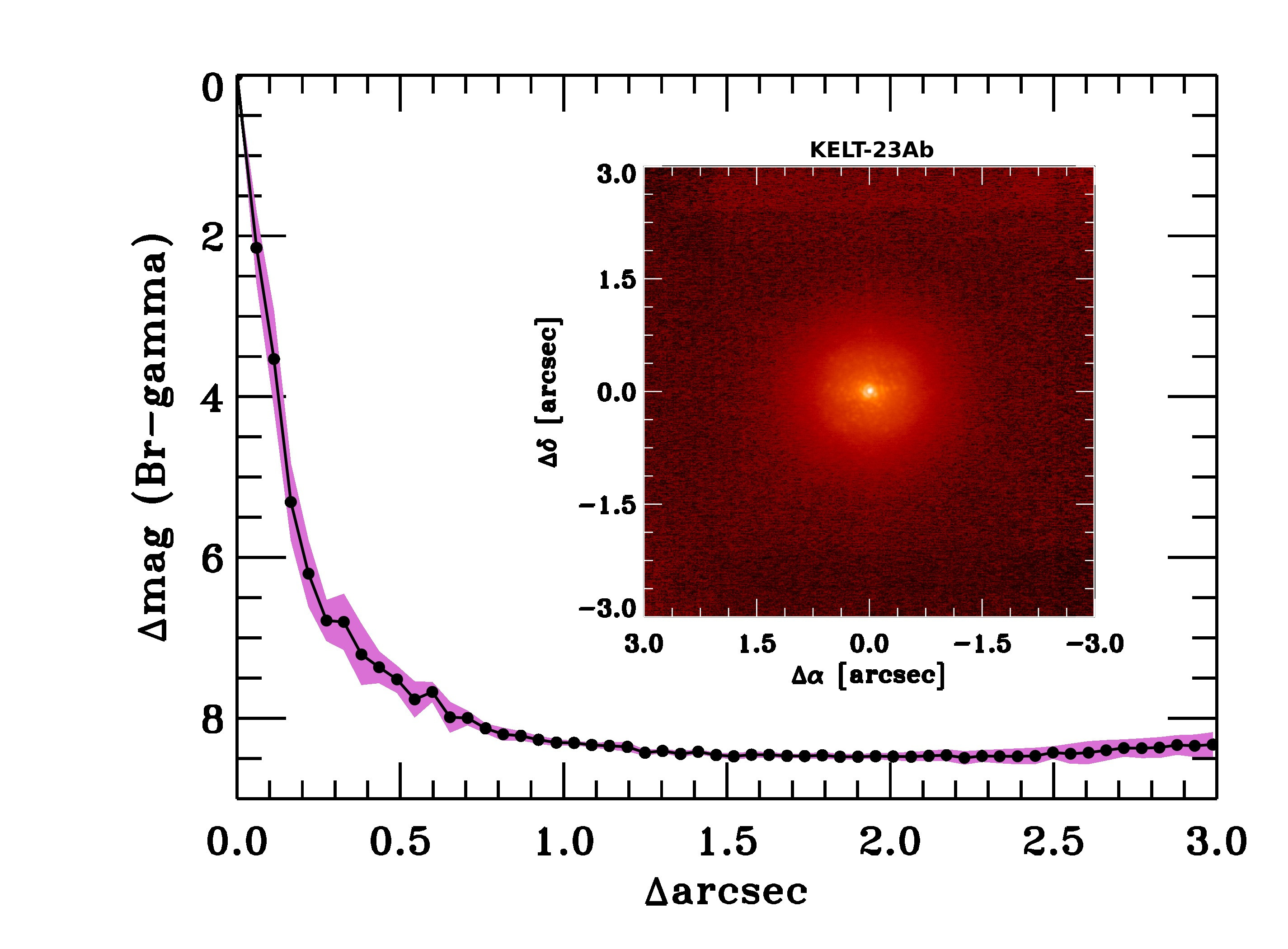}
    \caption{Companion sensitivity for the Keck adaptive optics imaging.  The black points represent the 5$\sigma$ limits and are separated in steps of 1 FWHM ($\sim 0.05$\arcsec); the purple represents the azimuthal dispersion (1$\sigma$) of the contrast determinations (see text). The inset image is of the primary target showing no addtional companions to within 3\arcsec\ of the target.}
    \label{fig:AO_fit}
\end{figure}

The observations were made in the narrow-band $Br-\gamma$ filter $(\lambda_o = 2.1686; \Delta\lambda = 0.0326\mu$m) with an integration time of 10 seconds with one coadd per frame for a total of 90 seconds on target.  The camera was in the narrow-angle mode with a full field of view of $\sim10\arcsec$ and a pixel scale of approximately $0.099442\arcsec$ per pixel. The Keck AO observations clearly detected a companion star 4.5$\arcsec$ to the southwest of the primary star which is the known star 2MASS J15283577+6621288 (Figure \ref{fig:AO_image}) and must be accounted for in the transit depth determinations for observations that do not resolve the stars \citep{Ciardi:2015}. No additional stellar companions were detected to within a resolution $\sim 0.05\arcsec$ FWHM (Figure \ref{fig:AO_fit}).

The sensitivities of the final combined AO image were determined by injecting simulated sources azimuthally around the primary target every $45^\circ $ at separations of integer multiples of the central source's FWHM \citep{Furlan:2017}. The brightness of each injected source was scaled until standard aperture photometry detected it with $5\sigma $ significance. The resulting brightness of the injected sources relative to the target set the contrast limits at that injection location. The final $5\sigma $ limit at each separation was determined from the average of all of the determined limits at that separation and the uncertainty on the limit was set by the rms dispersion of the azimuthal slices at a given radial distance. The sensitivity curve is shown in  Figure \ref{fig:AO_fit} along with an inset image zoomed to the primary target showing no other companion stars.

\begin{table}
	\centering
	\caption{Absolute RVs and Bisector Spans for KELT-23A\MakeLowercase{b}}
	\setlength{\tabcolsep}{2pt}
	\begin{tabular}{lrrrrl}
		\hline
		\hline
		\multicolumn{1}{l}{\bjdtdb} & \multicolumn{1}{c}{RV} 	& \multicolumn{1}{c}{$\sigma_{\rm RV}$} 	& \multicolumn{1}{c}{Bisector} &  \multicolumn{1}{c}{$\sigma_{\rm Bisector}$}	& Source \\
		& \multicolumn{1}{c}{($\rm m~s^{-1}$)} &\multicolumn{1}{c}{($\rm m~s^{-1}$)} &
		\multicolumn{1}{c}{($\rm m~s^{-1}$)} &\multicolumn{1}{c}{($\rm m~s^{-1}$)} \\
		\hline

        2458168.018618 & -162.7 & 27.3 & 20.0 & 21.8 & TRES \\
        2458184.972170 & 116.6 & 24.9 & 19.4 & 23.1 & TRES \\
        2458200.902902 & 75.4 & 19.1 & -20.8 & 14.2 & TRES \\
        2458201.932662 & -222.8 & 21.0 & -9.1 & 16.1 & TRES \\
        2458210.887269 & -200.7 & 20.1 & 48.1 & 13.9 & TRES \\
        2458211.973315 & 127.7 & 20.1 & 26.2 & 11.7 & TRES \\
        2458218.822284 & 105.3 & 23.6 & 6.7 & 16.3 & TRES \\
        2458227.783676 & 101.9 & 21.1 & -11.1 & 14.9 & TRES \\
        2458241.834480 & -56.7 & 20.2 & -22.3 & 14.2 & TRES \\
        2458243.807620 & 61.0 & 18.6 & 10.9 & 7.8 & TRES \\
        2458244.791342 & -174.6 & 20.1 & -6.9 & 11.0 & TRES \\
        2458245.744880 & 89.4 & 23.1 & -44.5 & 25.2 & TRES \\
        2458255.919545 & -217.4 & 16.3 & -19.1 & 12.1 & TRES \\
        2458263.868286 & 72.6 & 13.4 & -2.3 & 6.5 & TRES \\
        2458268.747169 & 16.4 & 14.6 & 4.0 & 12.5 & TRES \\
        2458269.736791 & -194.2 & 13.7 & -6.9 & 10.1 & TRES \\
        2458271.788423 & -204.9 & 14.9 & 8.7 & 8.9 & TRES \\
        2458273.749890 & -200.5 & 13.1 & -7.0 & 14.3 & TRES \\
        2458274.733876 & -19.2 &  12.3 & -5.0 & 8.6 & TRES \\
        2458275.708487 & -53.3 & 13.9 & -10.2 & 10.9 & TRES \\
        2458276.745735 & -113.8 & 13.0 & 21.2 & 17.7 & TRES \\
		2458167.067925 & 138.308 & 8.055 & 14.9 & 12.7 & APF \\
        2458207.851774 & 108.658 & 7.488 & 21.3 & 12.5 & APF \\
        2458210.942618 & -130.985 & 10.568 & 42.4 & 12.2 & APF \\
        2458222.808525 & 36.404 & 6.853 & 18.2 & 13.1 & APF \\
        2458230.805805 & -79.151 & 6.772 & 11.3 & 12.5 & APF \\
        2458233.858489 & -49.625 & 7.187 & 21.0 & 12.6 & APF \\
        2458275.762324 & -19.782 & 6.392 & 8.4 & 12.4 & APF \\
        2458283.751511 & 54.966 & 7.085 & 23.5 & 12.1 & APF \\
        2458285.739298 & -59.795 & 6.269 & 23.4 & 12.0 & APF \\
        2458288.721088 & 147.346 & 6.433 & 25.7 & 12.5 & APF \\
        2458289.708981 & -144.553 & 6.269 & 18.4 & 12.9 & APF \\

		\hline
		\hline
	\end{tabular}
    \label{tab:Spectra}
\end{table} 

\section{Host Star Properties}\label{sec:star_props}

\subsection{Spectral Analysis}\label{sec:spec_analysis}
Spectral analysis allows us to obtain initial estimates of several characteristics of KELT-23A. The TRES spectra were analyzed using the Spectral Parameter Classification (SPC) procedure outlined in \citet{Buchhave:2012}. SPC cross-correlates the spectra against a grid of stellar atmospheres made by \cite{Kurucz:1979, Kurucz:1992}. Allowing SPC to vary all parameters, we obtain a preliminary $\teff=\SPCteff$K, $\loggstar=\SPClogg$ (cgs), $\feh=\SPCfeh$, and $\vsini=\vsinistarval2$ km s$^{-1}$. The errors reported represent mean errors. The low $\vsini$ value indicates that the spectra are not significantly broadened by stellar rotation. The APF spectra were also analyzed and yielded a stellar effective temperature of $\APFteff$K and a [Fe/H] of $\APFfeh$. The larger quantity and SNR of the TRES spectra led us to accept the values obtained by the SPC over those obtained on the APF spectra.


\subsection{SED Analysis}\label{sec:SED}
Using KELT-23A's spectral energy distribution (SED)
together with the Gaia parallax, 
we were able to determine 
an empirical constraint on its radius. 
We use broadband photometry, shown in Table \ref{tbl:LitProps}, spanning 
the wavelength range 0.2--22
$\mu m$
from the GALEX NUV to the WISE4 passbands. 
We excluded the
2MASS $H$ and $K_S$ bands 
because they
contained the ``U" quality flag in the 2MASS catalog, meaning that the magnitude listed is an upper limit. 
 We adopted the $T_{\rm eff}$ and [Fe/H] from the spectrosopic analysis (see Section~\ref{sec:spec_analysis}), such that the one remaining free parameter of the fit is extinction, $A_V$.
 We use \cite{Kurucz:1979, Kurucz:1992} stellar atmosphere models to fit the SED, 
and the bolometric flux $F_{\rm bol}$ is obtained by numerically integrating the (unreddened) fitted SED model. 
The existence of a known companion within $4.5\arcsec$ (Section \ref{sec:ao}) indicates that the fluxes in the above bandpasses are blended, so we 
simultaneously fitted the companion star's SED by enforcing the same distance and extinction (see Section \ref{sec:companion}) but in this case fitting for $T_{\rm eff}$, and then subtracting the companion star's $F_{\rm bol}$ in order to obtain the true $F_{\rm bol}$ for KELT-23A alone. 

We obtain a good fit with reduced $\chi^2 = 4.4$, resulting in a best fit $A_V = 0.075 \pm 0.015$ and a (companion flux-corrected) $F_{\rm bol}$ for KELT-23A of $(2.056 \pm 0.071) \times 10^{-9}$ erg s$^{-1}$ cm$^{-2}$. With the Gaia DR2 parallax \citep[corrected for the systematic offset from][]{Stassun:2018}, we obtain
a preliminary stellar radius for KELT-23A of 0.959$\pm$0.016~$\rsun$.

\begin{figure}
	\includegraphics[width=0.75\linewidth, angle=90, clip=true]{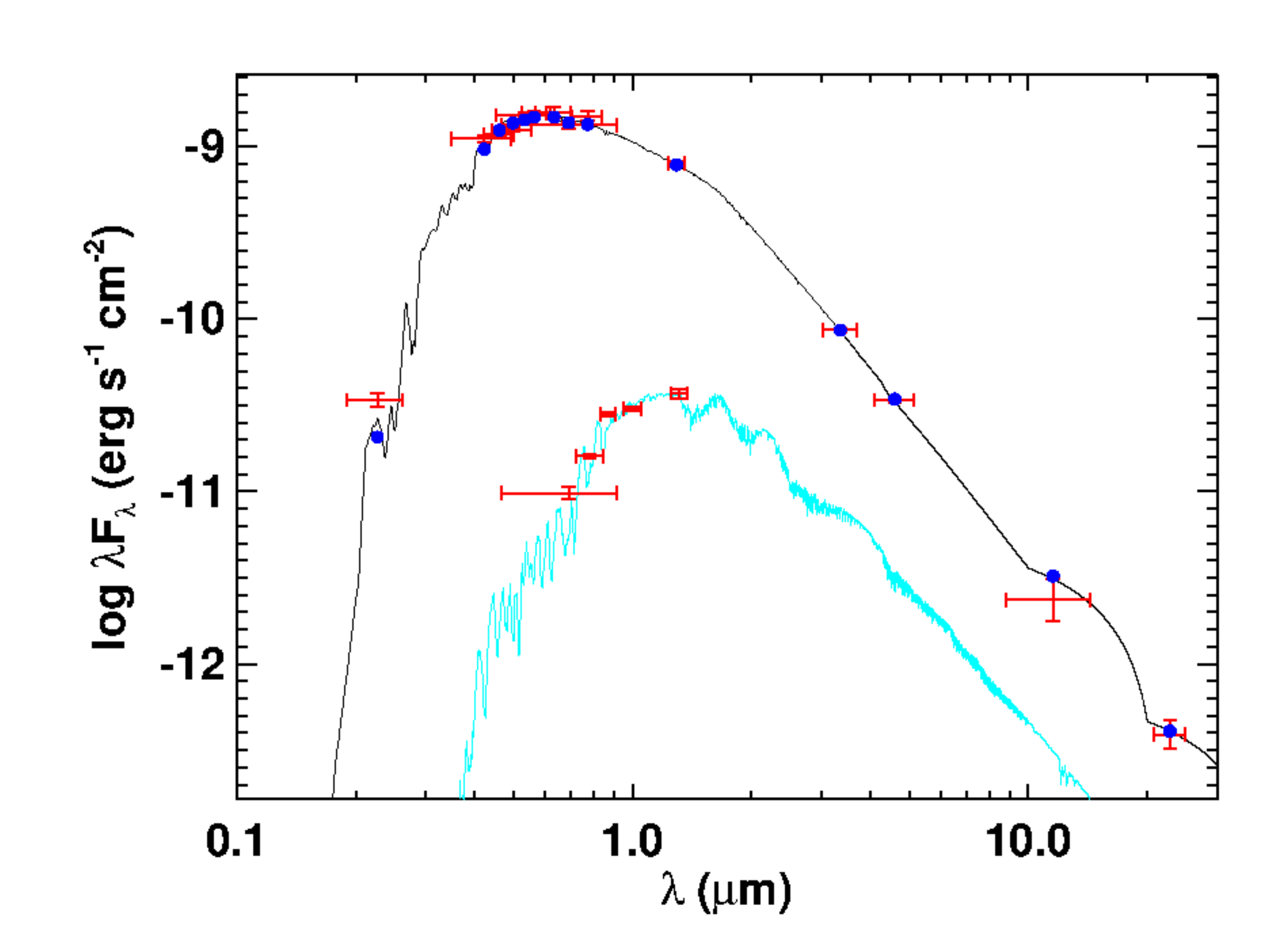}
	\caption{SED fits for KELT-23A (black line) and the fainter companion within 4.5$\arcsec$ (Section \ref{sec:ao}; light blue line). Red dots on KELT-23A's SED represent observed values for each bandpass in Table \ref{tbl:LitProps}. Horizontal errors represent the width of each bandpass and vertical errors represent the $1\sigma$ uncertainty of the fit. The blue dots represent the points of the model fit. Red dots on companion's SED result from our AO analysis, Gaia measurements, and Pan-STARRS measurements.}
	\label{fig:sed}
\end{figure}

\subsection{Stellar Models and Age}\label{sec:st_age}
We estimate the age of KELT-23A using the stellar parameters determined from the circular MESA Isochrones and Stellar Tracks (MIST; \citealt{Dotter:2016, Choi:2016, Paxton:2011, Paxton:2013, Paxton:2015}) model fit in Section \ref{sec:globalfit} and Table \ref{tab:KELT-23A}. We also determine an evolutionary track for KELT-23A, using a similar process to that in \cite{Siverd:2012}. In short, we find the intersection of the evolutionary track corresponding to the $\Mstar$ and $\feh$ with the best fitting isochrone for our $\teff$ and $\loggstar$. We find an age of $\agestarval$ Gyr, where the uncertainty only accounts for observational uncertainty in the parameters used in the fit. This does not account for systematic uncertainties or calibration uncertainties in the MIST model. We find that KELT-23A is a main-sequence G2V type star in the second half of its lifetime on the main-sequence, making KELT-23A a slightly older Solar-type star. 

The star's metallicity suggests that it could also be slightly sub-Solar. A more detailed analysis of individual elemental abundances of the host star is encouraged in order to search for deviations from the Solar abundance ratios. 

\begin{figure}
    \centering
    \includegraphics[width=\linewidth, trim={0.5cm 0cm 0cm 0.5cm }, clip]{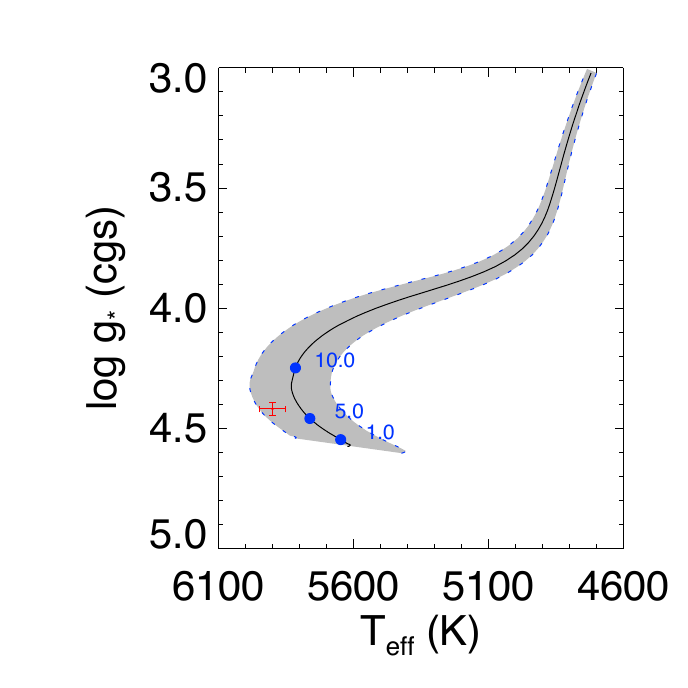}
    \caption{Age determination for KELT-23A. Plotted are KELT-23A with 1$\sigma$ uncertainties (red) and the best fitting MIST track (grey) with the shaded region representing 1$\sigma$ uncertainties on the host star's mass from our global fit. The $\Mstar$ and $\feh_0$ of the host star are determined in Section \ref{sec:globalfit} and are used in the HR analysis (Section \ref{sec:st_age}). Blue dots represent the position along the track at 1.0, 5.0, and 10.0 Gyr.}
    \label{fig:hrd}
\end{figure}

\section{Analysis and Results} \label{sec:analysis}

\subsection{EXOFASTv2 Global Fit}\label{sec:globalfit}

In order to gain a more complete understanding of the KELT-23A system's parameters, we simultaneously fit the ground based photometric observations from the KELT-Followup Network (See Figure \ref{fig:All_LCs}) and the radial velocities (See Figure \ref{fig:TRES_RVs}) from TRES using EXOFASTv2 \citep{Eastman:2013, Eastman:2017}. Within the global fit, the MIST stellar evolution models are simultaneously fit to constrain the stellar parameters. We place a Gaussian prior on the $\feh$ of -0.09$\pm$0.08 and $\teff$ of 5915$\pm$50 K from the SPC analysis of the TRES and APF spectra (See \S\ref{sec:spec_analysis}). We also place a Gaussian prior on $\Rstar$\ from the value given by the SED analysis (see \S\ref{sec:SED}, Figure \ref{fig:sed}). We perform two separate fits of KELT-23Ab, one where eccentricity is a free parameter and the other where eccentricity is enforced to be zero. The results of our two fits are shown in Table \ref{tab:KELT-23A} and \ref{tab:KELT-23A_other}.

\begin{table*}
 \scriptsize
\centering
\setlength\tabcolsep{1.5pt}
\caption{Median values and 68\% confidence intervals for the physical and orbital parameters of the KELT-23A system}
  \label{tab:KELT-18b_global_fit_properties}
  \begin{tabular}{lccccc}
  \hline
  \hline
   Parameter & Units & Value &\textbf{Adopted Value} \\
   & & (MIST eccentric) & \textbf{(MIST circular; $e$=0 fixed)}\\
  \hline
\multicolumn{2}{l}{Stellar Parameters:}&\smallskip\\
~~~~$M_*$\dotfill &Mass (\msun)\dotfill &$0.942^{+0.062}_{-0.055}$&$0.944^{+0.060}_{-0.054}$\\
~~~~$R_*$\dotfill &Radius (\rsun)\dotfill &$0.996\pm0.015$&$0.996\pm0.015$\\
~~~~$L_*$\dotfill &Luminosity (\lsun)\dotfill &$1.082^{+0.051}_{-0.049}$&$1.081^{+0.051}_{-0.048}$\\
~~~~$\rho_*$\dotfill &Density (cgs)\dotfill &$1.345^{+0.098}_{-0.087}$&$1.349^{+0.090}_{-0.083}$\\
~~~~$\log{g}$\dotfill &Surface gravity (cgs)\dotfill &$4.416^{+0.028}_{-0.027}$&$4.417^{+0.026}_{-0.025}$\\
~~~~$T_{\rm eff}$\dotfill &Effective Temperature (K)\dotfill &$5899\pm49$&$5899\pm49$\\
~~~~$[{\rm Fe/H}]$\dotfill &Metallicity (dex)\dotfill &$-0.106^{+0.078}_{-0.077}$&$-0.105^{+0.078}_{-0.077}$\\
~~~~$[{\rm Fe/H}]_{0}$\dotfill &Initial Metallicity \dotfill &$-0.072^{+0.070}_{-0.071}$&$-0.071\pm0.071$\\
~~~~$Age$\dotfill &Age (Gyr)\dotfill &$6.5^{+3.7}_{-3.3}$&$6.4^{+3.5}_{-3.2}$\\
~~~~$EEP$\dotfill &Equal Evolutionary Point \dotfill &$374^{+27}_{-32}$&$374^{+27}_{-30}$\\
~~~~$\dot{\gamma}$\dotfill &RV slope (m/s/day)\dotfill &$-0.196^{+0.079}_{-0.090}$&$-0.276^{+0.098}_{-0.11}$\\
\smallskip\\\multicolumn{2}{l}{Planetary Parameters:}&b&b\\
~~~~$P$\dotfill &Period (days)\dotfill &$2.255249\pm0.000012$&$2.255251^{+0.000011}_{-0.000012}$\\
~~~~$R_P$\dotfill &Radius (\rj)\dotfill &$1.323\pm0.025$&$1.323\pm0.025$\\
~~~~$T_C$\dotfill &Time of conjunction (\bjdtdb)\dotfill &$2458007.3184^{+0.0027}_{-0.0029}$&$2458007.3194^{+0.0027}_{-0.0028}$\\
~~~~$T_0$\dotfill &Optimal conjunction Time (\bjdtdb)\dotfill &$2458140.3781^{+0.0028}_{-0.0030}$&$2458140.3792^{+0.0027}_{-0.0029}$\\
~~~~$a$\dotfill &Semi-major axis (AU)\dotfill &$0.03300^{+0.00071}_{-0.00065}$&$0.03302^{+0.00068}_{-0.00064}$\\
~~~~$i$\dotfill &Inclination (Degrees)\dotfill &$85.37^{+0.31}_{-0.30}$&$85.37^{+0.31}_{-0.30}$\\
~~~~$e$\dotfill &Eccentricity \dotfill &$0.036^{+0.014}_{-0.013}$&---\\
~~~~$\omega_*$\dotfill &Argument of Periastron (Degrees)\dotfill &$3^{+36}_{-40}$&---\\
~~~~$T_{eq}$\dotfill &Equilibrium temperature (K)\dotfill &$1562\pm21$&$1561\pm20$\\
~~~~$M_P$\dotfill &Mass (\mj)\dotfill &$0.936^{+0.047}_{-0.043}$&$0.938^{+0.048}_{-0.044}$\\
~~~~$K$\dotfill &RV semi-amplitude (m/s)\dotfill &$150.5^{+3.8}_{-3.7}$&$150.6^{+4.4}_{-4.3}$\\
~~~~$logK$\dotfill &Log of RV semi-amplitude \dotfill &$2.178\pm0.011$&$2.178^{+0.012}_{-0.013}$\\
~~~~$R_P/R_*$\dotfill &Radius of planet in stellar radii \dotfill &$0.1365\pm0.0011$&$0.1365\pm0.0010$\\
~~~~$a/R_*$\dotfill &Semi-major axis in stellar radii \dotfill &$7.13^{+0.17}_{-0.16}$&$7.13^{+0.16}_{-0.15}$\\
~~~~$\delta$\dotfill &Transit depth (fraction)\dotfill &$0.01863^{+0.00031}_{-0.00030}$&$0.01864\pm0.00028$\\
~~~~$Depth$\dotfill &Flux decrement at mid transit \dotfill &$0.01863^{+0.00031}_{-0.00030}$&$0.01864\pm0.00028$\\
~~~~$\tau$\dotfill &Ingress/egress transit duration (days)\dotfill &$0.01706^{+0.0010}_{-0.00097}$&$0.01710^{+0.00089}_{-0.00085}$\\
~~~~$T_{14}$\dotfill &Total transit duration (days)\dotfill &$0.0992\pm0.0010$&$0.09921\pm0.00092$\\
~~~~$T_{FWHM}$\dotfill &FWHM transit duration (days)\dotfill &$0.08211^{+0.00060}_{-0.00061}$&$0.08208^{+0.00061}_{-0.00060}$\\
~~~~$b$\dotfill &Transit Impact parameter \dotfill &$0.575^{+0.028}_{-0.031}$&$0.576^{+0.024}_{-0.027}$\\
~~~~$b_S$\dotfill &Eclipse impact parameter \dotfill &$0.575^{+0.027}_{-0.028}$&---\\
~~~~$\tau_S$\dotfill &Ingress/egress eclipse duration (days)\dotfill &$0.0171^{+0.0011}_{-0.0010}$&---\\
~~~~$T_{S,14}$\dotfill &Total eclipse duration (days)\dotfill &$0.0994^{+0.0024}_{-0.0025}$&---\\
~~~~$T_{S,FWHM}$\dotfill &FWHM eclipse duration (days)\dotfill &$0.0822^{+0.0018}_{-0.0017}$&---\\
~~~~$\delta_{S,3.6\mu m}$\dotfill &Blackbody eclipse depth at 3.6$\mu$m (ppm)\dotfill &$1484\pm63$&$1482^{+64}_{-62}$\\
~~~~$\delta_{S,4.5\mu m}$\dotfill &Blackbody eclipse depth at 4.5$\mu$m (ppm)\dotfill &$1981\pm74$&$1980^{+76}_{-73}$\\
~~~~$\rho_P$\dotfill &Density (cgs)\dotfill &$0.502^{+0.038}_{-0.035}$&$0.503^{+0.039}_{-0.036}$\\
~~~~$logg_P$\dotfill &Surface gravity \dotfill &$3.123^{+0.027}_{-0.026}$&$3.124\pm0.027$\\
~~~~$\Theta$\dotfill &Safronov Number \dotfill &$0.0495^{+0.0016}_{-0.0015}$&$0.0496\pm0.0017$\\
~~~~$\fave$\dotfill &Incident Flux (\fluxcgs)\dotfill &$1.349^{+0.073}_{-0.070}$&$1.349^{+0.070}_{-0.067}$\\
~~~~$T_P$\dotfill &Time of Periastron (\bjdtdb)\dotfill &$2458006.80^{+0.23}_{-0.25}$&$2458007.3194^{+0.0027}_{-0.0028}$\\
~~~~$T_S$\dotfill &Time of eclipse (\bjdtdb)\dotfill &$2458006.234^{+0.015}_{-0.017}$&$2458008.4470^{+0.0027}_{-0.0028}$\\
~~~~$T_A$\dotfill &Time of Ascending Node (\bjdtdb)\dotfill &$2458006.776\pm0.018$&$2458006.7556^{+0.0027}_{-0.0028}$\\
~~~~$T_D$\dotfill &Time of Descending Node (\bjdtdb)\dotfill &$2458007.901^{+0.019}_{-0.017}$&$2458007.8832^{+0.0027}_{-0.0028}$\\
~~~~$ecos{\omega_*}$\dotfill & \dotfill &$0.030^{+0.011}_{-0.012}$&---\\
~~~~$esin{\omega_*}$\dotfill & \dotfill &$0.001\pm0.021$&---\\
~~~~$M_P\sin i$\dotfill &Minimum mass (\mj)\dotfill &$0.933^{+0.047}_{-0.043}$&$0.935^{+0.048}_{-0.044}$\\
~~~~$M_P/M_*$\dotfill &Mass ratio \dotfill &$0.000948\pm0.000030$&$0.000949\pm0.000033$\\
~~~~$d/R_*$\dotfill &Separation at mid transit \dotfill &$7.11^{+0.27}_{-0.25}$&$7.13^{+0.16}_{-0.15}$\\
~~~~$P_T$\dotfill &A priori non-grazing transit prob \dotfill &$0.1215^{+0.0045}_{-0.0044}$&$0.1210\pm0.0025$\\
~~~~$P_{T,G}$\dotfill &A priori transit prob \dotfill &$0.1599^{+0.0059}_{-0.0058}$&$0.1593\pm0.0035$\\
~~~~$P_S$\dotfill &A priori non-grazing eclipse prob \dotfill &$0.1211\pm0.0029$&---\\
~~~~$P_{S,G}$\dotfill &A priori eclipse prob \dotfill &$0.1594^{+0.0042}_{-0.0041}$&---\\
\hline
\label{tab:KELT-23}
 \end{tabular}
\end{table*}

\begin{table*}[ht!]
\label{tab:other}
\scriptsize
\centering
\setlength\tabcolsep{1.5pt}
\caption{Median values and 68\% confidence intervals for the additional parameters of KELT-23A from EXOFASTv2 }
  \begin{tabular}{lccccccccccc}
  \hline
  \hline
\smallskip\\\multicolumn{2}{l}{Wavelength Parameters:}&\\
&linear limb-darkening &quadratic limb-darkening &Dilution \\
&~~~~$u_{1}$\dotfill &~~~~$u_{2}$\dotfill &~~~~$A_D$\dotfill\\
$B$&$0.612\pm0.050$&$0.202\pm0.050$&$0.0013336\pm0.0000067$\\
$I$&$0.268\pm0.034$&$0.292\pm0.034$&$0.002565\pm0.000013$\\
$R$&$0.342\pm0.047$&$0.289\pm0.048$&$0.003504\pm0.000018$\\
$g^{\prime}$&$0.512^{+0.047}_{-0.046}$&$0.227\pm0.047$&$0.0014614\pm0.0000073$\\
$i^{\prime}$&$0.315^{+0.050}_{-0.049}$&$0.312\pm0.049$&$0.009836\pm0.000049$\\
$r^{\prime}$&$0.374\pm0.045$&$0.300\pm0.047$&$0.003357\pm0.000017$\\
$z^{\prime}$&$0.200\pm0.034$&$0.258\pm0.034$&$0.02120\pm0.00011$\\
$V$&$0.448\pm0.036$&$0.300\pm0.035$&$0.002566\pm0.000013$\\
\smallskip\\\multicolumn{2}{l}{Telescope Parameters:}&\multicolumn{2}{c}{Mist Eccentric}&\multicolumn{2}{c}{Mist Circular}\\
&&APF&TRES&APF&TRES\smallskip\\
~~~~$\gamma_{\rm rel}$\dotfill &Relative RV Offset (m/s)\dotfill &$7.2^{+4.0}_{-3.8}$&$-59.7^{+4.8}_{-4.5}$&$7.6^{+5.5}_{-5.2}$&$-56.1^{+4.9}_{-4.8}$\\
~~~~$\sigma_J$\dotfill &RV Jitter (m/s)\dotfill &$9.1^{+5.5}_{-4.3}$&$5.4^{+7.4}_{-5.4}$&$14.4^{+6.3}_{-4.3}$&$8.4^{+6.5}_{-8.4}$\\
~~~~$\sigma_J^2$\dotfill &RV Jitter Variance \dotfill &$82^{+130}_{-60}$&$28^{+130}_{-78}$&$210^{+220}_{-110}$&$70^{+150}_{-89}$\\
\hline
\smallskip\\\multicolumn{2}{l}{Transit Parameters (MIST circular Fit):}&\\
Observation & Added Variance & TTV (days)& Baseline flux & detrending coeff & detrending coeff & detrending coeff \\
&$\sigma^{2}$&(days)& $F_0$ & $C_{0}$& $C_{1}$& $C_{2}$\\
WCO UT 2018-01-26 ($I$)&$-0.00000126^{+0.00000071}_{-0.00000062}$&$0.0088^{+0.0029}_{-0.0028}$&$0.99953\pm0.00022$&$-0.00120^{+0.00062}_{-0.00061}$&$-0.00034^{+0.00047}_{-0.00046}$&$-0.00117^{+0.00062}_{-0.00063}$\\
PvdK UT 2018-01-26 ($r^{\prime}$)&$0.00000673^{+0.00000077}_{-0.00000068}$&$0.0076^{+0.0029}_{-0.0028}$&$1.00025\pm0.00021$&$-0.00067\pm0.00060$&$-0.00089\pm0.00068$&$-0.00061\pm0.00061$\\
DEMONEXT UT 2018-02-03 ($g^{\prime}$)&$0.0000231^{+0.0000019}_{-0.0000018}$&$0.0073^{+0.0029}_{-0.0028}$&$1.00001\pm0.00028$&$-0.00017\pm0.00064$&$0.00079^{+0.00063}_{-0.00064}$&---\\
KeplerCam UT 2018-02-04 ($i^{\prime}$)&$0.0000132^{+0.0000015}_{-0.0000013}$&$0.0062^{+0.0029}_{-0.0028}$&$0.99974\pm0.00031$&$0.00022\pm0.00049$&---&---\\
astroLAB UT 2018-02-17 ($I$)&$0.0000002^{+0.0000061}_{-0.0000051}$&$0.0059^{+0.0029}_{-0.0028}$&$1.00163^{+0.00064}_{-0.00063}$&$-0.0014\pm0.0016$&$-0.0012\pm0.0016$&---\\
PvdK UT 2018-03-10 ($z^{\prime}$)&$0.00000648^{+0.0000011}_{-0.00000099}$&$0.0062^{+0.0030}_{-0.0029}$&$0.99924\pm0.00025$&$0.00081\pm0.00072$&$0.00061\pm0.00073$&---\\
ASP UT 2018-03-19 ($B$)&$0.0000218^{+0.0000043}_{-0.0000039}$&$0.0099^{+0.0032}_{-0.0030}$&$0.99946^{+0.00051}_{-0.00052}$&$-0.0011\pm0.0014$&$-0.0025\pm0.0016$&$-0.0018\pm0.0014$\\
ASP UT 2018-03-19 ($R$)&$0.0000126^{+0.0000019}_{-0.0000017}$&$0.0106^{+0.0030}_{-0.0029}$&$1.00072\pm0.00027$&$-0.00143^{+0.00073}_{-0.00072}$&$-0.00108\pm0.00073$&---\\
WCO UT 2018-03-19 ($z^{\prime}$)&$0.0000008^{+0.0000019}_{-0.0000016}$&$0.0138^{+0.0034}_{-0.0032}$&$1.00005^{+0.00061}_{-0.00062}$&$0.00071^{+0.00099}_{-0.0010}$&---&---\\
SOTES UT 2018-05-31 ($V$)&$0.0000055^{+0.0000025}_{-0.0000020}$&$0.0131^{+0.0034}_{-0.0033}$&$1.00059\pm0.00041$&$-0.00183^{+0.00095}_{-0.00096}$&$0.00035\pm0.00097$&---\\
KUO UT 2018-07-03 ($V$)&$0.00000294^{+0.00000084}_{-0.00000074}$&$0.0121^{+0.0034}_{-0.0032}$&$0.99865\pm0.00032$&$-0.00298^{+0.00094}_{-0.00095}$&$-0.00147^{+0.00096}_{-0.00097}$&---\\
\hline
  \hline
\label{tab:KELT-23_other}
 \end{tabular}
\end{table*}

\subsection{UVW Space Motion}\label{sec:UVW}
We computed the three-dimensional space motion of KELT-23A to characterize it kinematically in a Galactic context. The systemic velocity was measured to be $\absolutervval$ km s$^{-1}$~(Section \ref{sec:spectroscopy}), which is within $1\sigma$ of the value reported by Gaia DR2 ($RV_{Gaia}=-14.891\pm0.379$; \citealt{Katz:2018}), and the proper motion of the system was found to be $\mu_{\alpha}=0.434\pm0.039$ mas~yr$^{-1}$ and $\mu_{\delta}=-12.217\pm0.041$ mas~yr$^{-1}$ (Table \ref{tbl:LitProps}). We adopt the distance value derived from the Gaia parallax (Table \ref{tbl:LitProps}), and the standard of rest values from \cite{Coskunoglu:2011}. We calculate the space motion to be ($U$, $V$, $W$) = ($17.32\pm0.03$, $0.95\pm0.07$, $-0.90\pm0.07$) km s$^{-1}$. When compared with the distributions in \cite{Bensby:2003}, we find a 99.4\% probability that KELT-23A is a member of the thin disk population in the Galaxy. KELT-23A's near-Solar metallicity and relatively low velocities are consistent with the age derived in Section \ref{sec:st_age}.

\subsection{Transit Timing Variation Results}\label{sec:ttvs}
Transit timing variations (TTVs) can be evidence of an unobserved planetary companion gravitationally perturbing the observed planet \citep{Agol:2005, Holman:2005}. Hot Jupiters seldom show significant TTVs because they rarely have companions close to their mean motion resonances \citep{Steffen:2012}. However, there is at least one example of a Hot Jupiter that exhibits TTVs due to a nearby companion, specifically WASP-47b, which was discovered by \cite{Hellier:2012} and whose TTVs due to a nearby Neptune-mass companion were discovered in K2 data by \cite{Becker:2015}. Therefore, it is important that we continue to search for Hot Jupiters with detectable TTVs, as these rare systems can provide rich and precious information about the emplacement mechanisms of hot Jupiters.

\begin{figure}
    \includegraphics[width=\linewidth]{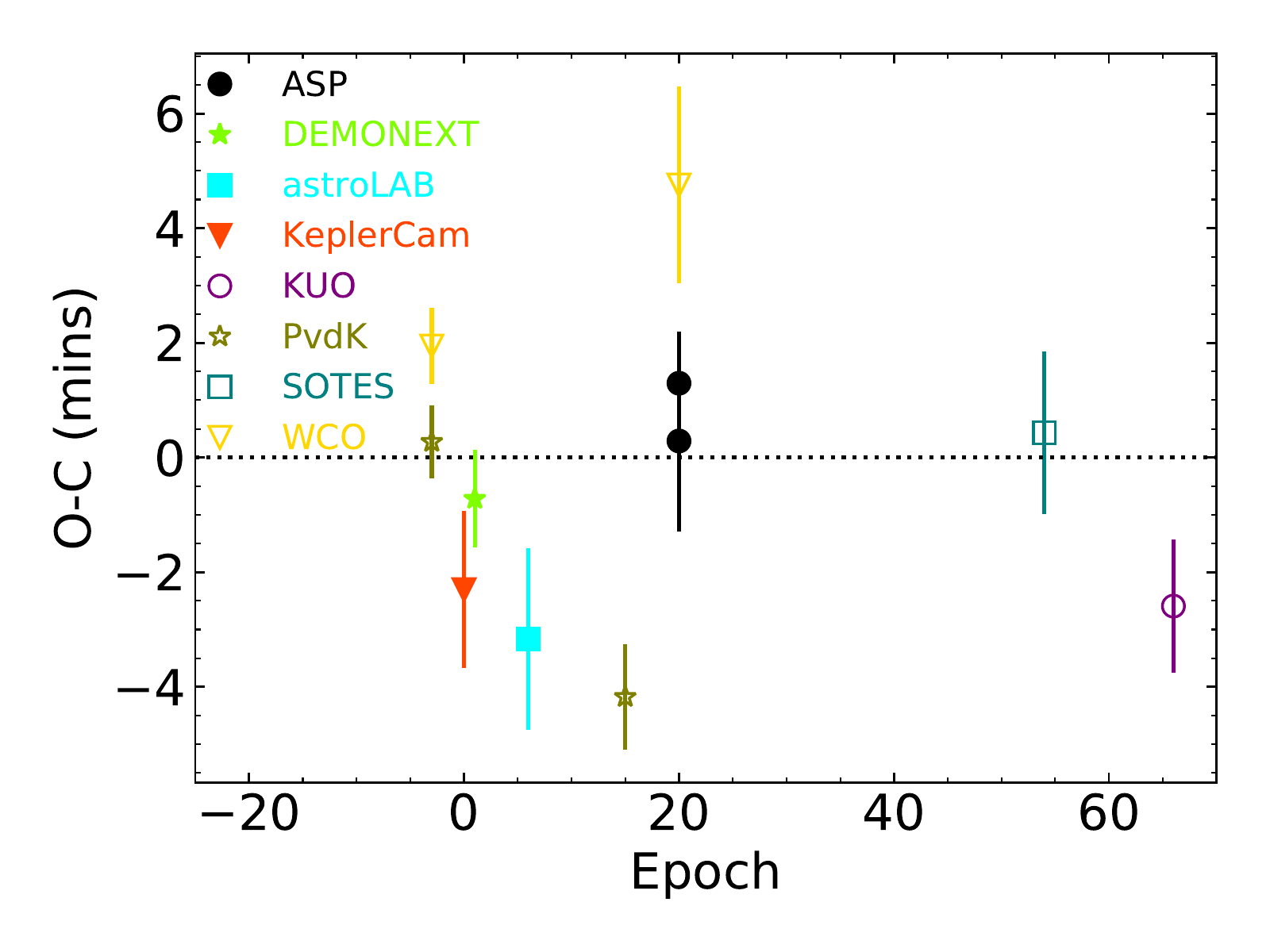}
    \caption{The transit time residuals for KELT-23Ab using the global fit ephemeris determined in Section \ref{sec:globalfit}. The data are listed in Table \ref{tab:TTV}.}
    \label{fig:TTVs}
\end{figure}

To analyze any possible TTVs we first converted each observatory's photometric time stamps to \bjdtdb~\citep{Eastman:2010} before the global fit in Section \ref{sec:globalfit}. The accuracy of the time stamps is assured by the synchronization to a standard clock by each observatory during observations. This synchronization assures that any uncertainty in $T_0$ is a result of the global fit. We then plotted the O-C residuals between the observed transit times and the calculated transit times using the ephemeris determined in Section \ref{sec:globalfit} (Table \ref{tab:KELT-23A}). The O-C residuals are in Table \ref{tab:TTV} and plotted in Figure \ref{fig:TTVs}. Similar to the KELT-14 TTV analysis \citep{Rodriguez:2016}, it would appear that there are significant TTVs. However, observatory-dependent factors such as seeing conditions and astrophysical red-noise \citep{Carter:2009} can induce discrepancies in the observed transit mid-times. Due to these factors, we do not claim that the observed TTVs are astrophysical in nature and conclude that they are likely spurious and a result of systematics. Continued monitoring, especially at the precision that will be reached by TESS, will unveil any true TTVs.

\begin{table}
	\centering
	\caption{Transit Times for KELT-23Ab}
	\begin{tabular}{lccccc}
		\hline
		\hline
		\multicolumn{1}{l}{Epoch} &
		\multicolumn{1}{c}{$T_0$} &
		\multicolumn{1}{c}{$\sigma_{T_0}$} & 
		\multicolumn{1}{c}{O-C} & 
		\multicolumn{1}{c}{O-C} &
		Telescope \\
		&
		\multicolumn{1}{c}{(\bjdtdb)} & 
		\multicolumn{1}{c}{(s)} &
		\multicolumn{1}{c}{(s)} &
		\multicolumn{1}{c}{($\sigma_{T_0})$} & \\
		\hline
		-3 & 2458144.8984 & 40 & 116 & 2.9 & WCO \\
		-3 & 2458144.89724 & 38 & 16 & 0.4 & PvdK \\
		1 & 2458153.91793 & 51 & -43 & -0.8 & DEMONEXT \\
		0 & 2458153.91681 & 82 & -138 & -1.7 & KeplerCam \\
		6 & 2458167.4484 & 95 & -190 & -2.0 & astroLAB \\
		15 & 2458187.74583 & 55 & -250 & -4.5 & PvdK \\
		20 & 2458196.77035 & 95 & 17 & 0.2 & ASP \\
		20 & 2458196.77106 & 54 & 77 & -1.4 & ASP \\
		20 & 2458196.7735 & 103 & 285 & 2.7 & WCO \\
		54 & 2458273.45249 & 85 & 25 & 0.3 & SOTES \\
		66 & 2458302.76997 & 70 & -155 & -2.2 & KUO \\
		\hline
		\hline
	\end{tabular}
    \label{tab:TTV}
\end{table}

\subsection{Evidence for a Wide Binary Star System} \label{sec:companion}
As with some other recently discovered hot-Jupiter hosts, such as KELT-18 \citep{McLeod:2017}, there is evidence that KELT-23A is a member of a wide binary star system.  The nearby star described in Section \ref{sec:ao} is most likely gravitationally bound to KELT-23A. The agreement of the parallaxes of both components within their sub-one percent uncertainties, and the consistency of their relative proper motion with their expected orbital motion, lead us to conclude that the nearby star is bound to KELT-23A.

The nearby star's Gaia DR2 parallax \citep[also corrected for the systematic offset from][]{Stassun:2018} is $7.977 \pm 0.052$ mas and places the star at a distance of $125.36\pm0.82$ pc, equal to KELT-23A's Gaia-derived distance to within the small uncertainty.  At that distance, with an apparent angular separation of 4.5$\arcsec$, the projected (minimum) distance between the two stellar components would be 570 AU.

The nearby star's Gaia DR2 proper motions of $\mu_{\alpha}=1.567\pm0.092$ mas~yr$^{-1}$ and $\mu_{\delta}=-11.902\pm0.107$ mas~yr$^{-1}$ are close to KELT-23A's (Table \ref{tbl:LitProps}) and give the two stars a relative proper motion between them of just 1.2 mas yr$^{-1}$.  This corresponds to a relative tangential velocity of 0.72 km s$^{-1}$, which could be produced by orbital motion.  For illustration, a 0.25 M$_{\odot}$ companion at 570 AU from KELT-23A would be in a 12,000 yr orbit.  KELT-23A would have a (circular) orbital speed of 0.28 km s$^{-1}$ and the companion's orbital speed would be 1.12 km s$^{-1}$.  A 0.5 M$_{\odot}$ companion would orbit with a period of 11,000 yr giving KELT-23A and its companion (circular) orbital speeds of 0.52 km s$^{-1}$ and 1.0 km s$^{-1}$, respectively.  The SED analysis (Section \ref{sec:SED}) produces relative fluxes for the companion (Table \ref{tab:RelPhotComp}) that are consistent with a $\sim$ 0.3 M$_{\odot}$ star.

\begin{table}
	\centering
	\caption{Relative fluxes for the companion star, determined by the SED analysis (Section \ref{sec:SED}).}
	\begin{tabular}{cc|cc}
		\hline
		\hline
		Filter & $F_B / F_A$ & Filter & $F_B / F_A$ \\
		\hline
		$B$ & 0.00134 & $g'$ & 0.00146 \\
        $V$ & 0.00257 & $r'$ & 0.00337 \\
        $R$ & 0.00352 & $i'$ & 0.00993 \\
        $I$ & 0.01282 & $z'$ & 0.02166 \\
		\hline
		\hline
	\end{tabular}
    \label{tab:RelPhotComp}
\end{table}

The presence of KELT-23A's distant companion adds it to a family of systems that are informing the developing theories about the formation of hot Jupiters.  One way to drive a gas giant planet inward toward its host star is through eccentric orbital interactions with a distant companion star via Kozai-Lidov oscillations \citep{Fabrycky:2007}.  The planet's orbit would later circularize due to tidal friction as the planet approaches the host star.

\section{False Positive Analysis} \label{sec:false_pos}
KELT-23Ab underwent the same rigorous vetting process used for all KELT planet candidates. We first obtain KELT-FUN photometry of the candidate (Section \ref{sec:phot}). Follow-up observers have instruments capable of observing the target at faster cadences with greater photometric accuracy and greater angular resolution than KELT. In addition, follow-up photometric observations ideally span a large wavelength range in order to check for wavelength dependent transit depths. Our observations span the $B$ - $z'$ range and show no wavelength dependent anomalies. All of our observed light curves are well fit with our global planetary model. The model consists of an opaque circular planet of constant achromatic radius on a Keplerian orbit. (Section \ref{sec:globalfit}). 

An indication that our follow-up TRES spectra are contaminated with a close stellar partner would be double absorption lines. Our 11 TRES spectra and 15 APF spectra show no double absorption lines. We find no evidence for a stellar partner blended in the spectra, signifying that the periodic RV signal arises from KELT-23A.

To rule out any other sources of periodic radial velocity anomalies we analyzed the bisector spans of each spectrum \citep{Buchhave:2010}, found in Table \ref{tab:Spectra}. Between the RVs and the BSs, we calculated a Spearman correlation of 0.14 and a p-value of 0.4285, indicating that there is no correlation between the bisector spans and the radial velocities. We conclude that the observed periodic radial velocity changes are due to the reflex motion of the star caused by the transiting companion.

As a final check, the AO images were analyzed for a companion within 5$\arcsec$ of KELT-23A. There exists a close stellar companion within $4.5\arcsec$ of KELT-23A, which was previously known. This companion was not resolved by the photometry and measures were taken to deblend the fluxes by examining the flux ratios between KELT-23A and the stellar companion in the Johnson-Cousins and SDSS filters. Through the previously mentioned checks, we conclude that the photometric anomalies and spectroscopic RV signal are best described by a planet orbiting KELT-23A.

\section{Discussion} \label{sec:discussion}

\subsection{Tidal Evolution and Irradiation History} \label{sec:Irradiation}

We simulate the past and future evolution of KELT-23Ab's orbit using the parameters derived in Section \ref{sec:globalfit} as boundary conditions. We use the {\it POET} code for this simulation \citep{Penev:2014} under several assumptions: a circular orbit, no perturbing companion planets, and a constant phase lag (constant tidal quality factor). The simulation analyzes the change in semi-major axis over time due to the tidal forces exchanged by the star and planet. The simulation also analyzes the change in incident flux received by the planet over time. This change in incident flux is a result of both the changing semi-major axis and host star luminosity as it evolves. Tidal force strengths are parameterized through the tidal dissipation parameter, Q$'_*$. This parameter is defined as the quotient of the tidal quality factor (Q$_*$) and the Love number (k$_2$). We test values of Q$'_*$ ranging from 10$^5$ to 10$^8$ stepping up only the order of magnitude. This large range in tidal dissipation factor allows us to probe a large range of timescales for energy dissipation. These timescales correspond to the large range of proposed mechanisms for tidal dissipation.

\begin{figure}
    \centering
    \includegraphics[width=\linewidth, clip]{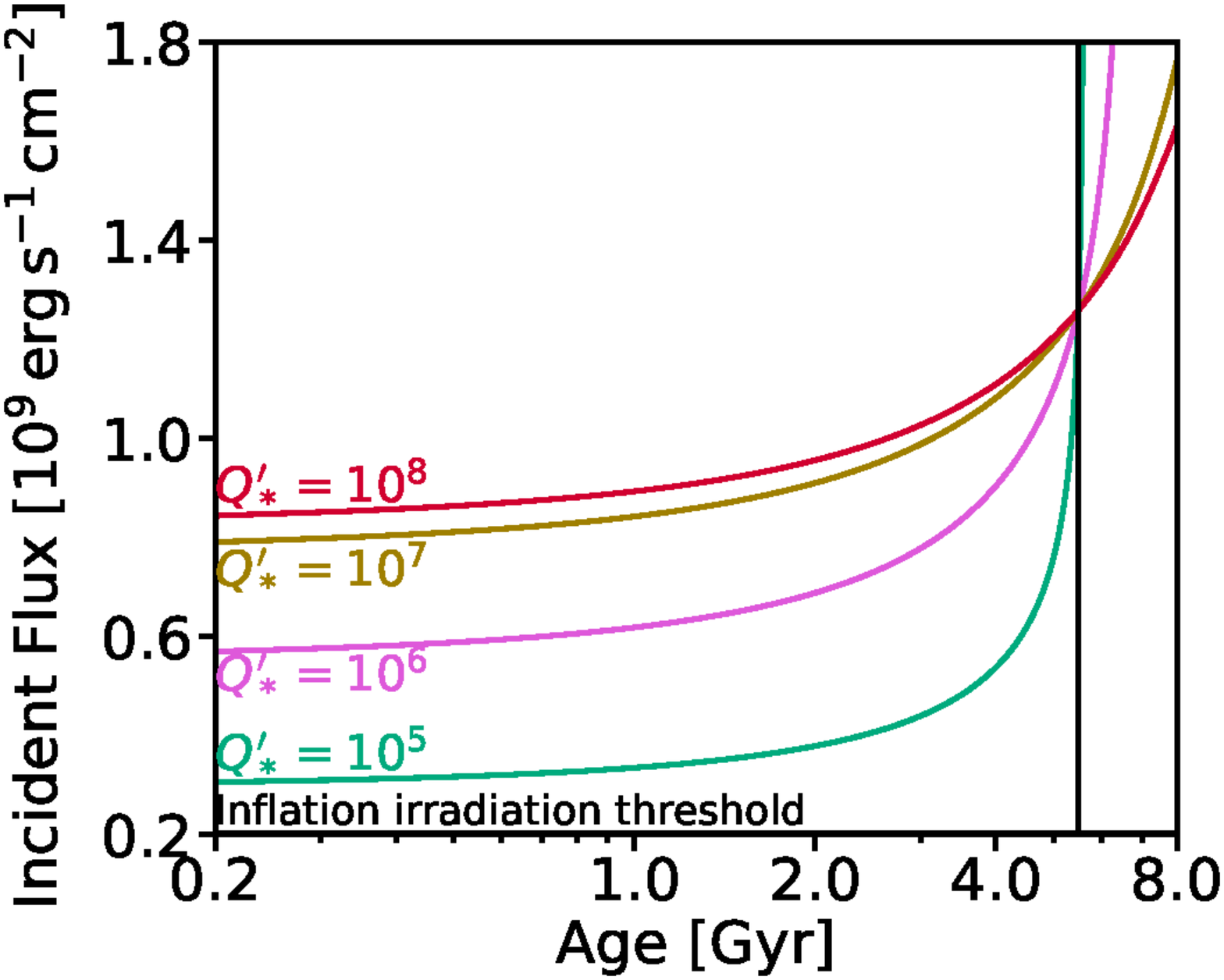}
    \includegraphics[width=\linewidth, clip]{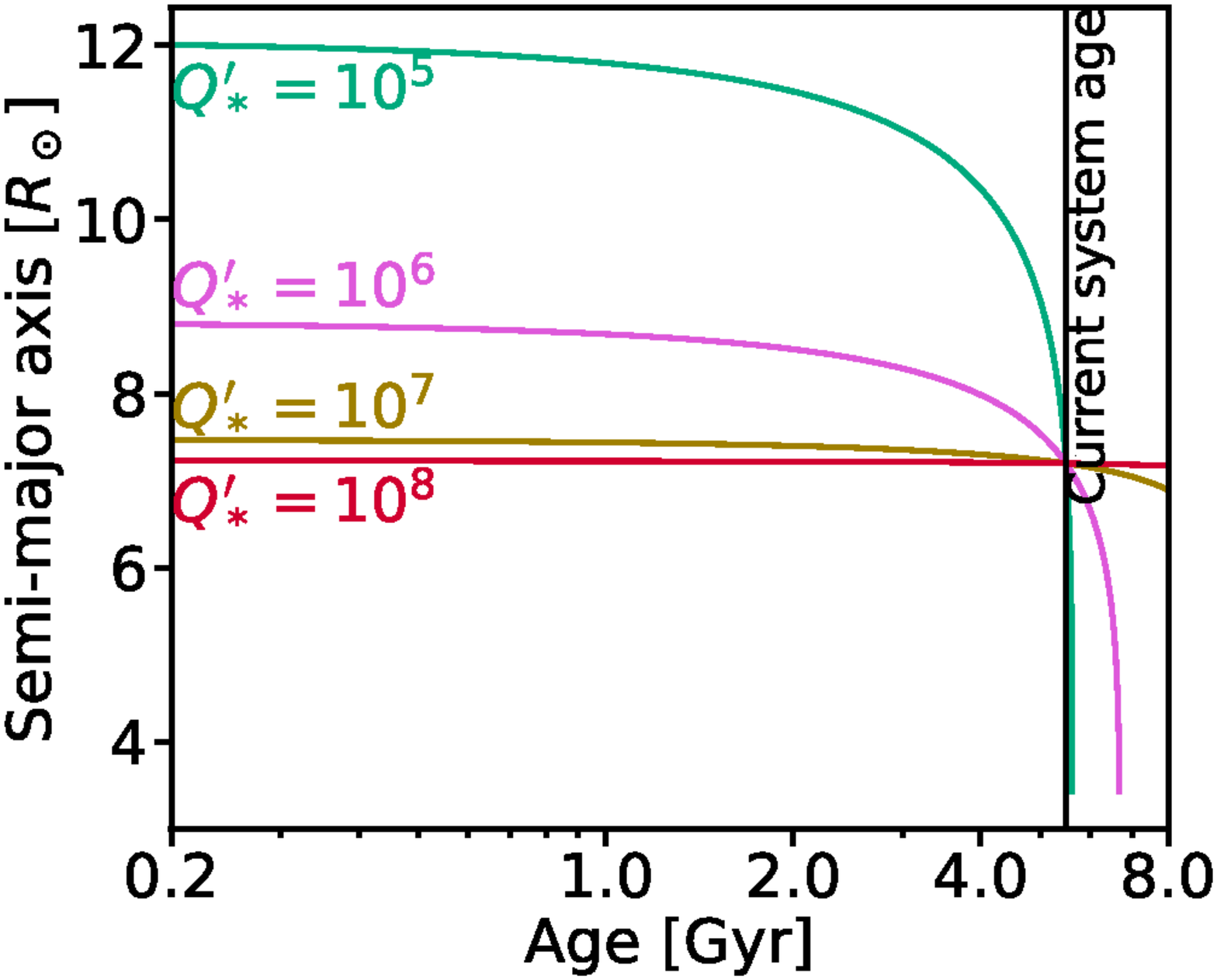}
    \caption{{\it Top}: Irradiation history of KELT-23Ab modeled for a range of stellar tidal quality factors, Q$'_*$. KELT-23Ab is shown to be above the inflation irradiation threshold ($2\times10^8$ erg s$^{-1}$ cm$^{-2}$; \citealt{Demory:2011}) throughout its history. {\it Bottom}: Change in semi-major axis of KELT-23Ab over time modeled from several Q$'_*$ values}
    \label{fig:Kalo}
\end{figure}

If tides have been the primary cause of changing orbital properties, then KELT-23Ab has been above the inflation irradiation threshold, as defined in \cite{Demory:2011} ($2\times10^8$ erg s$^{-1}$ cm$^{-2}$), for its entire history, regardless of the Q$'_*$ chosen. Currently, KELT-23Ab receives an incident flux about 6 times larger than the Demory \& Seagar inflation threshold (Figure \ref{fig:Kalo}; Upper panel). KELT-23Ab's large radius, $R_{\textrm P}=\rplanetval$ $R_{\textrm J}$, is almost certainly due to its long history of high stellar irradiation.

The future of KELT-23Ab is highly dependent on the value of Q$'_*$. A large value (10$^8$) will keep the planet in orbit for several Gyrs. A small value (10$^5$) will quickly cause the planet to spiral into its host star, as seen in the bottom panel of Figure \ref{fig:Kalo}.

If, in addition to the orbital properties, the spin of the star is known, then it is possible to constrain $Q_{*}^{'}$, following the procedure of \cite{Penev:2018}. Briefly, the procedure relies on comparing the observed spin of the star to the one predicted by gyrochronology and assuming that any excess spin is due to tidal dissipation transferring orbital angular momentum to the star. Not having a measurement of the stellar spin, we can instead simply evaluate the frequency-dependent tidal dissipation found in \cite{Penev:2018} at the tidal period of this system, resulting in an expected value of log($Q_{*}^{'}$)$=5.8$. If this truly is the case then KELT-23Ab is expected to spiral into its host within the next Gyr or so.

While KELT-23Ab is not an extreme case of star-planet tidal interactions like WASP-18b \citep{Hellier:2009}, the differences in host star properties, mainly the prevalence of a larger surface convection zone on KELT-23A, could potentially lead to observable TTVs in the system over a long time period.  Nonetheless, tidal interactions within the KELT-23Ab system are almost certainly strong enough to conspire with potential Kozai-Lidov oscillations to produce a hot Jupiter, if the nearby star is indeed bound to KELT-23A.

 
\subsection{Future Observations}

The KELT-23A system presents a unique opportunity for follow-up from {\it TESS} and the {\it JWST}. KELT-23Ab has the second-highest ecliptic latitude ($\beta=75\arcdeg.112$) of any transiting exoplanet, second only to Qatar-1b \citep{Alsubai:2011}. The system's high ecliptic latitude makes it visible to the {\it JWST} 61\% of the year.  {\it JWST} will be able to characterize planetary atmospheres through the transmission spectroscopy method \citep{Beichman:2014} and KELT-23Ab will be a well-suited candidate for extensive atmospheric studies due to its high equilibrium temperature ($\teq=\teqplanetvalshort$ K) and relatively low surface gravity ($\loggp=\loggplanetvalshort$). It is also possible to probe the dayside thermal emission of the planet by obtaining infrared photometry near secondary eclipse, as described by \cite{Rodriguez:2016} in reference to KELT-14b. 

The system will also receive multi-sector coverage from {\it TESS}, which can be used to determine the transit ephemeris with extremely high precision and detect any TTVs too small to observe from the ground. {\it TESS}'s photometric precision would allow for the detection of an Earth-sized companion at an orbital period of several days. {\it TESS} will observe many consecutive transits that can be used to study details of the system's transit morphology. 

{\it TESS}'s relatively long observing baseline for KELT-23A will also allow detailed characterization of the planet using the {\it TESS} light curve alone. KELT-23A should be observed by {\it TESS} in three sectors during Year 2 of its mission. Given KELT-23A's {\it TESS} bandpass magnitude of $T=9.739$, we estimate that {\it TESS} should achieve a per-point photometric precision of 180 ppm hr$^{-1/2}$ on KELT-23A. Over the three 26-day sectors, {\it TESS} will observe $\sim34.5$ orbital periods of KELT-23Ab. Phase-folding the {\it TESS} data on the period implies a binned light curve precision of $\sim30$ ppm hr$^{-1/2}$. Additionally, TESS's high photometric precision could detect microvariability of the host star due to stellar oscillations \citep{Campante:2016}.

We use the \citet{Price:2014} expressions for the analytic precision estimates on the transit observables given finite exposure times to estimate the precision with which {\it TESS} can measure the time of transit center $T_C$, the ingress/egress durations $\tau$, and the transit and eclipse depths. With the aforementioned per-point precision and estimated number of observable transits, {\it TESS} should measure $T_C$ to $10^{-4}$ days, approximately 9 seconds, at both the two-minute or 30-minute cadences (a factor of 10 improvement over the current precision), the ingress/egress duration $\tau$ to fractional uncertainties of $2\%$ and $6\%$ at each cadence, and the transit depth $\delta$ to 0.5\% at both cadences. Following \citet{Stevens:2018}, the improvement in the precision on $\tau$ at the two-minute cadence would improve the stellar density precision to $\approx 3\%$ (from the current $6\%$) in the limit that the ingress/egress duration is the dominant source of uncertainty.

We calculate the anticipated secondary eclipse depth due to reflected light via Equation (39) of \citet{Winn:2010}. The depth is $4\times10^{-4}A_{\lambda}$, where $A_{\lambda}$ is the geometric albedo. For even a very large albedo, $A_{\lambda} = 0.3$, the fractional uncertainties on the reflected light eclipse depths are 71\% and 79\% at two-minute and 30-minute cadence.

Assuming that the temperature of KELT-23Ab is equal to the equilibrium temperature, $T_{\rm eq} = 1534$K, as well as perfect heat redistribution in its atmosphere, we calculate the expected secondary eclipse depth from KELT-23Ab's thermal emission to be 5 mmag in the {\it TESS} passband; the fractional uncertainty is calculated to be 2\% at both {\it TESS} cadences, so thermal emission should be readily detectable.

Using this temperature, the results from our global fit, and Equation (35) of \citet{Winn:2010}, we calculate a scale height of $3\times10^{-3}$\rj\ and an expected transmission signal of $10^{-4}$, as is expected for a typical hot Jupiter. 

We estimated the amplitudes of three components of the phase curve of KELT-23Ab (Doppler beaming, ellipsoidal variations, and reflection) using the equations presented in \cite{Faigler:2011}, \cite{Shporer:2011}, and \cite{Shporer:2018}, and our measured parameters of KELT-23Ab. The amplitude of each phase curve component, however, depends upon a coefficient $\alpha$ which is of order unity \citep[see Eqns.~1-3 of][]{Shporer:2011}. We assumed that the beaming coefficient $\alpha_{\mathrm{beam}}=1$, and calculated the ellipsoidal and reflection coefficients per \cite{Shporer:2018} and \cite{Faigler:2011}, respectively, using the linear limb darkening coefficient from the $i'$ band (as an approximation to the {\it TESS} bandpass) and estimated albedo, respectively. This resulted in $\alpha_{\mathrm{ellip}}=1.08$, and $\alpha_{\mathrm{refl}}=0.3$ ($3.0$) if the planetary geometric albedo is $0.03$ ($0.3$). These imply phase curve amplitudes of $A_{\mathrm{beam}}\sim2$ ppm, $A_{\mathrm{ellip}}\sim3$ ppm, and $A_{\mathrm{refl}}\sim11$ ($110$) ppm. Comparing these to the estimated phase curve precision, this suggests that the beaming and ellipsoidal components will be undetectable, while the reflected light may be detectable, depending upon the albedo. To first order, however, the amplitude of the thermal phase curve should be similar to the secondary eclipse depth (assuming a large day-night temperature contrast). Our estimated {\it TESS} secondary eclipse depth of 4.9 mmag is much larger than the reflected light component, suggesting that the {\it TESS} phase curve will be dominated by thermal emission from the planet.


\section{Conclusion} \label{sec:conclusion}
KELT-23Ab is an inflated hot Jupiter with an orbital period of $\periodvalshort$ days hosted by a G2V solar analog. The star-planet tidal interactions suggest that KELT-23Ab could be consumed by its host in about 1 Gyr. 
Systems like KELT-23A pose important questions as to why some Solar-type stars host hot Jupiters and others do not. {\it TESS} will provide many transits that will be used to search for more subtle lightcurve effects in the secondary eclipse from the planet's reflection and thermal emission. We calculate that the thermal emission from the planet will be comparable to the secondary eclipse depth and will be easily detectable in the {\it TESS} phase curve. KELT-23Ab's close proximity to the {\it JWST} CVZ means it will be visible for 61\% of the year. This long-term visibility makes KELT-23Ab an excellent follow-up candidate for atmospheric characterization from the {\it JWST}. 

\acknowledgements
This project makes use of data from the KELT survey, including support from The Ohio State University, Vanderbilt University, and Lehigh University, along with the KELT follow-up collaboration.
Work performed by J.E.R. was supported by the Harvard Future Faculty Leaders Postdoctoral fellowship.
D.J.S. and B.S.G. were partially supported by NSF CAREER Grant AST-1056524.
Work by S.V.Jr. is supported by the National Science Foundation Graduate Research Fellowship under Grant No. DGE-1343012.
Work by G.Z. is provided by NASA through Hubble Fellowship grant HST-HF2-51402.001-A awarded by the Space Telescope Science Institute, which is operated by the Association of Universities for Research in Astronomy, Inc., for NASA, under contract NAS 5-26555.
This paper includes data taken at The McDonald Observatory of The University of Texas at Austin.
This work has made use of NASA's Astrophysics Data System, the Extrasolar Planet Encyclopedia, the NASA Exoplanet Archive, the SIMBAD database operated at CDS, Strasbourg, France, and the VizieR catalogue access tool, CDS, Strasbourg, France.  We make use of Filtergraph, an online data visualization tool developed at Vanderbilt University through the Vanderbilt Initiative in Data-intensive Astrophysics (VIDA).
We also used data products from the Widefield Infrared Survey Explorer, which is a joint project of the University of California, Los Angeles; the Jet Propulsion Laboratory/California Institute of Technology, which is funded by the National Aeronautics and Space Administration; the Two Micron All Sky Survey, which is a joint project of the University of Massachusetts and the Infrared Processing and Analysis Center/California Institute of Technology, funded by the National Aeronautics and Space Administration and the National Science Foundation; and the European Space Agency (ESA) mission {\it Gaia} (\url{http://www.cosmos.esa.int/gaia}), processed by the {\it Gaia} Data Processing and Analysis Consortium (DPAC, \url{http://www.cosmos.esa.int/web/gaia/dpac/consortium}). Funding for the DPAC has been provided by national institutions, in particular the institutions participating in the {\it Gaia} Multilateral Agreement. Any opinions, findings, and conclusions or recommendations expressed are those of the author and do not necessarily reflect the views of the National Science Foundation. This material is based upon work supported by the National Science Foundation under Grant No. 1559487. KP acknowledges support from NASA grants 80NSSC18K1009 and NNX17AB94G. We thank Alex Jensen for observations that contributed to this work. JL acknowledges support from FAPESP (grant 2017/23731-1). We also thank the anonymous referee for valuable comments that improved this manuscript. DJ acknowledges support from the Carol and Ray Neag Undergraduate Research Fund.

\software{
		  \texttt{numpy} \citep{numpy},
          \texttt{matplotlib} \citep{matplotlib},
          \texttt{scipy}, \citep{SciPy},
          \texttt{astropy}, \citep{astropy}
          }

\clearpage
\bibliographystyle{aasjournal}
\bibliography{KELT-23b}

\end{document}